\documentclass[12pt]{iopart}

\usepackage{graphicx}
\usepackage{subfigure}
\usepackage[colorlinks=true,linktocpage=true,linkcolor=blue,citecolor=blue]{hyperref}
\usepackage[usenames,dvipsnames]{color}
\usepackage{placeins}
\usepackage{bm} 
\definecolor{darkblue}{RGB}{0,0,160}

\newcommand{\checked}[1]{}

\def\ba{\begin{eqnarray}}
\def\ea{\end{eqnarray}}
\def\lb{\label}
\def\nn{\nonumber \\}

\def\d{\delta}
\def\D{\Delta}

\def\o{\omega}

\def\rr{\rightarrow}

\def\e{\eta}

\def\t{\tau_f}

\def\p{\perp}

\def\sp{\;\;\;\;}
\def\th{\theta}

\begin{document}

\title[BWM description of HBT radii]{Blast-wave model description of the Hanbury-Brown--Twiss radii in $pp$ collisions \\ at LHC energies}

\author{Andrzej Bialas}
\address{M. Smoluchowski Institute of Physics, Jagellonian University, PL-30-059~Krakow, Poland}
\ead{bialas@th.if.uj.edu.pl}

\author{Wojciech Florkowski}
\address{ $^1$Institute of Physics, Jan Kochanowski University, PL-25406~Kielce, Poland}
\address{ $^2$The H. Niewodnicza\'nski Institute of Nuclear Physics, Polish Academy of Sciences, PL-31342 Krak\'ow, Poland} 
\ead{Wojciech.Florkowski@ifj.edu.pl}

\author{Kacper Zalewski}
\address{ $^1$M. Smoluchowski Institute of Physics, Jagellonian University, PL-30-059~Krakow, Poland}
\address{ $^2$The H. Niewodnicza\'nski Institute of Nuclear Physics, Polish Academy of Sciences, PL-31342 Krak\'ow, Poland} 
\ead{zalewski@th.if.uj.edu.pl}

\begin{abstract}
The blast wave model is applied to  the recent data  on HBT radii in $pp$ collisions, measured by the ALICE collaboration. A reasonable description of data is obtained for a rather low temperature of the kinetic freeze-out, $T\simeq$ 100 MeV, and the transverse profile corresponding to the emission from a shell of a fairly small width $2 \d \sim 1.5$~fm.  The size and the life-time of the produced system are determined  for various multiplicities of the produced particles.
\end{abstract}


\vspace{2pc}
\maketitle

\section{Introduction}  
\label{sect:intro}

Recently, the ALICE collaboration has presented an impressive collection of data on  the Hanbury-Brown--Twiss (HBT) radii measured in $pp$ collisions at the 7 TeV center-of mass energy~\cite{Aamodt:2011kd}. In the present paper we discuss to what extent this data is consistent with the blast-wave model \cite{Siemens:1978pb,Schnedermann:1993ws,
Florkowski:2004tn} which has been used  in analyses of the HBT correlations in relativistic heavy-ion collisions~\cite{Retiere:2003kf,Kisiel:2006is,
Florkowski:2010zz}. We note that the blast-wave model,  originally introduced in \cite{Siemens:1978pb}, was inspired by the results of the hydrodynamic description of the hadron production processes. It was later adapted to ultra-relativistic energies in \cite{Schnedermann:1993ws}, for a short review see also \cite{Florkowski:2004tn}. 

The agreement of the data with the blast-wave model predictions suggests that the produced matter exhibits thermal features such as local equilibration and hydrodynamic flow. As a matter of fact, at high energies, such as those presently available at the LHC, the final state hadron multiplicities are large, and a thermodynamic/hydrodynamic description  of hadron production may possibly be valid even in more elementary hadron+hadron and hadron+nucleus collisions, e.g., see Ref.~\cite{Ferroni:2011fh,Bozek:2013ska,Shapoval:2013jca}. Recently, the blast wave model has been used in this context to analyze high-multiplicity $pp$ collisions at the LHC \cite{Ghosh:2014eqa}. The authors of \cite{Ghosh:2014eqa} found indications for the strong transverse radial flow in such events. 

In the present paper we consider  the blast wave model featuring a boost-invariant, azimuthally symmetric  fluid expanding in the transverse direction according to the Hubble law \cite{Chojnacki:2004ec}.
We also assume  that the momentum distribution of the particles emitted from the fluid element with the four-velocity $u$ at freeze-out is given  by the Boltzmann formula
  \ba
   e^{-\beta E^*}=e^{-\beta p^\mu u_\mu},
  \ea
where $E^*$ is the energy of the emitted particle in the fluid element rest frame, $p^\mu$ is the particle four-momentum, and $T=1/\beta$ is the temperature of the system.
   
The main conclusion of this work is that the blast-wave model can indeed account for the vast collection of the ALICE data \cite{Aamodt:2011kd}. However, a suitably chosen transverse profile for the distribution of matter in the transverse plane should be used in order to describe the data well. This profile corresponds to  a shell of  radius $R$ and  width~$2\d$ (with $\d < R$). Interestingly, such a shape helps to reproduce correctly the ratio $R_{\rm out}/R_{\rm side}$ of the two HBT radii measured at large values of the transverse momentum of the pion pair. The permanent problems with a correct reproduction of this ratio are known as the {\it HBT puzzle}. In heavy-ion physics these problems may be eliminated in practice if several improvements/modifications are done in the standard hydrodynamic codes   \cite{Broniowski:2008vp,Pratt:2008qv}. In this context, our present finding offers yet another hint on a possible solution of the HBT puzzle.
 
\medskip
The paper is organized as follows: In the next Section we define the model by introducing the source function based on the Cooper-Frye formula and Hubble-like expansion of the fluid. The momentum distribution of particles, the HBT correlation functions and the HBT radii are discussed in Sections 3 and 4. The results of the data analysis are described in Sections 5 and 6. The results are summarized in the last Section. The Appendix contains the tables and figures where the model results are compared with the data.

\section{The Source function }
\label{sect:CF}

Our starting point  is the formula for the source/emission function
\begin{eqnarray}
S(x,p) = \int d\Sigma_\mu(x) \, p^\mu f(x)  e^{-\beta p^\mu u_\mu(x)} .
\label{Sxp1}
\end{eqnarray}
Here $x$ and $p$ are the spacetime position and four-momentum of the emitted particle (which we anticipate to be a pion) and $d\Sigma_\mu(x)$ is an element of the freeze-out hypersurface which we take in the form
\begin{eqnarray}
d\Sigma_\mu(x) &=& S_0  \sigma_\mu(x)
\, \delta(\tau_f - \tau) d^4x =S_0  \sigma_\mu(x)
\, \delta(\tau_f - \tau)\tau d\tau d\e d^2r \,,
\label{dSigma}
\end{eqnarray}
where the variables $\tau$ and $\eta$ are the longitudinal proper time and the space-time rapidity %
\begin{eqnarray}
t=\tau\cosh\e, \sp z=\tau \sinh\e.
\end{eqnarray}
In a similar way, we define the particle radial distance  from the beam axis and the azimuthal angle in the transverse plane
\begin{eqnarray}
x=r\cos\phi, \sp y=r\sin\phi.
\end{eqnarray}
The four-vector $\sigma^\mu=\sigma^\mu(x)$ defines the space-time orientation of an element of the freeze-out hypersurface 
\begin{eqnarray}
\sigma^\mu = \left(\cosh\eta,0,0,\sinh\eta\right).
\label{sigma}
\end{eqnarray}
The function $f(x)$ in (\ref{Sxp1}) describes the distribution of particles in space. Following Ref.~\cite{Bialas:2014gca} we assume that $f(x)$ depends only on the transverse radius $r$. Below we argue that the appropriate choice of the distribution $f(r)$ is crucial for reproducing the experimental results.

Since the system is boost-invariant and cylindrically symmetric, the four-velocity  $u=u(x)$ has the form \cite{Florkowski:2010zz}
\begin{eqnarray}
u &=&   
\left(\cosh\eta \cosh\th, 
\sinh\theta \cos\phi, 
\sinh\theta \sin\phi, 
\sinh\eta \cosh\theta \right).
\label{u}
\end{eqnarray}
In addition, we assume that the transverse rapidity of the fluid element at freeze-out~$\theta(r)$ and its position $r$ are related by the condition of the radial Hubble-like  flow \cite{Chojnacki:2004ec}. This leads to the expressions
\begin{eqnarray}
\sinh\theta = \omega r, \quad
\cosh\theta = \sqrt{1+\omega^2 r^2},
\label{hub}
\end{eqnarray} 
where $\omega$ is the parameter controlling the strength of the transverse flow.

The particle four-momentum is parameterized in the standard way in terms of rapidity, $y$, transverse momentum, $p_\perp$, transverse mass, $m_\perp$, and the azimuthal angle in the transverse plane, $\phi_p$,
\begin{eqnarray}
p = \left(m_\perp \cosh y, p_\perp \cos\phi_p, p_\perp \sin\phi_p, m_\perp \sinh y \right).
\label{p}
\end{eqnarray}
The scalar product of $p$ and $u$ is
\begin{eqnarray}
p \cdot u = m_\perp \cosh(y-\eta) \cosh\theta - p_\perp \cos(\phi_p-\phi) \sinh\theta.
\label{pu}
\end{eqnarray}
This form is used in the thermal Boltzmann distribution. In a similar way we obtain the factor $p \cdot \sigma$ needed to define the element of the freeze-out hypersurface
\begin{eqnarray}
p \cdot \sigma = m_\perp  \cosh(y-\eta).
\label{psigma}
\end{eqnarray}
The form of (\ref{psigma}) follows directly from (\ref{p}) and (\ref{pu}). Other forms are also possible here if one assumes different freeze-out conditions. Using (\ref{dSigma}) and (\ref{sigma}) we obtain the most popular version of the blast-wave model.

\section{Momentum distribution and the HBT correlation functions}
\label{sect:mom+corr}

The integral of the source function $S(x,p)$ over the space-time coordinates gives the momentum distribution
\ba
\frac{dN}{dy d^2p_\perp} = W(p)= \int d^4x \,S(p,x).
\ea
The calculation  starting from Eq.~(\ref{Sxp1}) leads to the expression \cite{Schnedermann:1993ws,Florkowski:2004tn} \footnote{From now on we shall omit all constant factors in the source function, since its normalization is irrelevant  for the problems we discuss in this paper.}
\ba
W(p_\p)=  m_\p \int r dr  f(r)   K_1(U) I_0(V), \lb{wpt}
\ea
where $K_1$ and $I_0$ are the modified Bessel functions and 
\ba
U=\beta m_\p \cosh\theta, \sp V=\beta p_\p \sinh\theta.
\label{UandV}
\ea

Assuming that one can neglect correlations between the produced particles,  the distribution of two identical bosons can be expressed in terms of the Fourier transform of the source function \cite{Bialas:2013oza}
\ba
W(p_1,p_2)=W(p_1)W(p_2) +|H(P,Q)|^2
\ea
with
\begin{eqnarray} 
H(P,Q)&=&\int d^4x e^{iQ \cdot x} S(x,P).
\label{hpq}
\end{eqnarray}
Here $Q=p_1-p_2$ and $\vec{P}=(\vec{p}_1+\vec{p}_2)/2$. The time-component of the four-vector $P$ is not uniquely defined. We take  $P_0=\sqrt{m^2+|\vec{P}|^2}$ \cite{Pratt:2008qv}.
\medskip
We shall  work in the so-called LCMS system in which $P_z=0$, i.e., $p_{1z}=-p_{2z}$. In this reference frame the substitution $p\rr P$ in the source function $S(x,p)$ is simply realized by the change $m_\p\rr \sqrt{P_0^2-P_z^2}= P_0$. Starting directly from (\ref{hpq}) we find
\begin{eqnarray}
H(P,Q) &=& P_0   
\int r\, dr f(r) \int d\phi 
 \int d\eta  \cosh\eta e^{-U \cosh\eta + V \cos\phi - i \Phi }
\label{FT1}
\end{eqnarray}
where now $U=\beta P_0\cosh\theta$ and $V=\beta P_\perp \sinh\theta$, and the phase $\Phi$ is given by the formula
\begin{eqnarray}
\Phi = -Q_0 t +Q_z z + Q_x x+ Q_y y. \label{Phi}
\end{eqnarray}
The phase $\Phi$ depends on the  relative direction of $\vec{P}=(P_\perp,0,0)$ and $\vec{Q}$. Following the standard approach \cite{LL}, we  consider three regimes: $long$, $side$ and $out$~\footnote{Note that we use the notation $P_\perp$ for the transverse momentum of a pion pair instead of $k_T$ used in Ref.~\cite{Aamodt:2011kd}.}. It was shown in~\cite{Bialas:2014gca} that $H_d(P_\perp,q)$ can be explicitly expressed as integrals involving  Bessel functions. We have
\ba
H_{\rm long}=P_0 \int rdrf(r) I_0(V) U K_1(U_l)/U_l, \sp U_l=\sqrt{U^2+Q_z^2\t^2}, \lb{hl}
\ea
\ba 
H_{\rm side}= P_0 \int rdrf(r) I_0(V_s) K_1(U), \sp V_s=\sqrt{V^2-Q_y^2r^2}. \lb{hs}
\ea 
If $V^2<Q_y^2r^2$, $V_s$ is  imaginary and the function $I_0(V_s)$ should  be replaced by $J_0(|V_s|)$. In the $out$ direction we have
\ba
H_{\rm out} = P_0 \int rdrf(r) I_0(V+iQ_xr) K_1(U-iQ_0\t ). \lb{ho}
\ea

\section {The HBT radii}
\label{sect:HBTradii}

Experiments usually measure the correlation function defined as   
\ba
C(p_1,p_2)\equiv \frac{W(p_1,p_2)}{W(p_1)W(p_2)} -1=\frac{|H(P,Q)|^2}{W(p_1)W(p_2)}.
\ea
Each HBT radius is obtained from a gaussian fit to the correlation function for the  direction $long$, or $side$, or $out$:
\ba
C(p_1,p_2)=e^{-R^2_{\rm HBT}q^2},
\ea
with $q$ being the component of the vector $\vec{Q}$ in the analyzed direction. This means  that the radii can be evaluated analytically as the logarithmic derivatives of the correlation functions at $q=0$
\ba
R^2_{\rm HBT}=-\frac{d \log[C(p_1,p_2)]}{dq^2} |_{q=0}. \lb{rform}
\ea

Using this definition and the formulae from the previous Section one  obtains the expressions for the $R^2_{\rm HBT}$ in all three directions, in the form of integrals involving the modified Bessel functions. They were given explicitly in \cite{Bialas:2014gca} and, as they are rather lengthy, we shall not repeat them here.

\section{Comparison with data}

The  HBT radii were measured by the ALICE collaboration for 6 intervals  of pair transverse momentum and 8 intervals of multiplicity. This means that, at each multiplicity interval, there are 18 numbers to be explained. Our aim is thus to check if these 
18 experimental numbers can be accounted for by the model, and --- where possible --- to determine the relevant  physical parameters. 

Two parameters, the temperature $T=1/\beta$ and $\omega$, (responsible for the transverse flow, c.f. Eq.~(\ref{hub}) )  reflect the dynamics of the produced  system, whereas the transverse profile $f(r)$  describes its geometry.

In our analysis we have assumed that the temperature is  constant (i.e., its value is fixed on the freeze-out hypersurface and independent of the multiplicity class). It turned out that an acceptable $\chi^2$ can be obtained only if $T$ does not exceed 120~MeV --- calculations done with higher values of $T$,  not presented here, result in much worse values of $\chi^2$ (exceeding the number of degrees of freedom). In the final analysis presented below we use  $T=100$ MeV. 

\subsection{The transverse profile}

For the transverse profile we took a two-parameter function
\ba
f(r)\sim e^{-(r-R)^2/\d^2}
\ea
(normalization is irrelevant for our purposes), i.e., we consider emission from a shell of radius $R$ and  width $2\d$. Note that this form includes, as a special case ($R=0$), the gaussian profile, sometimes used in description of the heavy-ion data. We found, however, that to obtain a good description of the ALICE $pp$ data it is necessary to keep $R >\d$. 

\subsection {The transverse momentum}

Since the model must be consistent with the general features of data, it is  necessary to demand the agreement with the measured (average) transverse momentum.  This condition implies an additional  relation between the parameters of the model.  In order to implement this condition, we observe that, as seen from (\ref{wpt}), the distribution of transverse momentum depends on  three parameters: $T$, $ \o R$ (controlling the transverse flow) and $\D\equiv \delta/R$ (describing the shape of the transverse profile). Demanding that the average transverse momentum resulting from (\ref{wpt}) agrees with the data of Ref.~\cite{cmspt},  one finds a relation between $\o R$ and $\D$ (at  a fixed value of $T$). Thus, effectively, we are left with three parameters: $\t$, $R$ and $\d=\D R$ for the description of the HBT radii (note that we fit only the average transverse momentum of pions).  

The measurements by the CMS collaboration \cite{cmspt} give, approximately, 
\ba
\langle P_\p \rangle \approx [400+ 2.5(N_c-10)] \hbox{ MeV},   \lb{ptav}
\ea
where $N_c$, the number of  charged particles, ranges between 10 and 50. Using (\ref{ptav}) as  input, one can find numerically the relation between $\o R$ and $\D$. For $T=100$ MeV it can be approximated by the formula
\ba
\o R=a_0(N_c)+a_1(N_c)\D+a_2(N_c)\D^2+a_3(N_c)\D^3
\label{omegaRx}
\ea
with the following coefficients:

\ba
      a_0=0.695+0.00785\;N_c-0.0000075\;N_c^2,\nn
      a_1=-0.385-0.00395\;N_c-0.0000075\;N_c^2,\nn
      a_2=0.0868+0.00062\;N_c+0.0000085\;N_c^2,\nn
      a_3=-0.00312+0.0000865\;N_c-0.00000272\;N_c^2.
 \label{as}     
\ea

\section {Description of the HBT radii} 

To reduce further the number of independent model parameters we have accepted the simple idea of selecting  $\d$, the half-width of the "shell" from which particles are emitted, to be constant, independent of  multiplicity (and thus also of the size of the system).  As shown below, the 7 TeV ALICE data are consistent with this assumption.  It should be emphasized, however, that data do not restrict substantially $\d$, particularly at low multiplicities.  It is  thus not excluded that  the condition $\d=$ const may be challenged by more precise future measurements.

\begin{figure}[t]
\begin{center}
\includegraphics[scale=0.33]{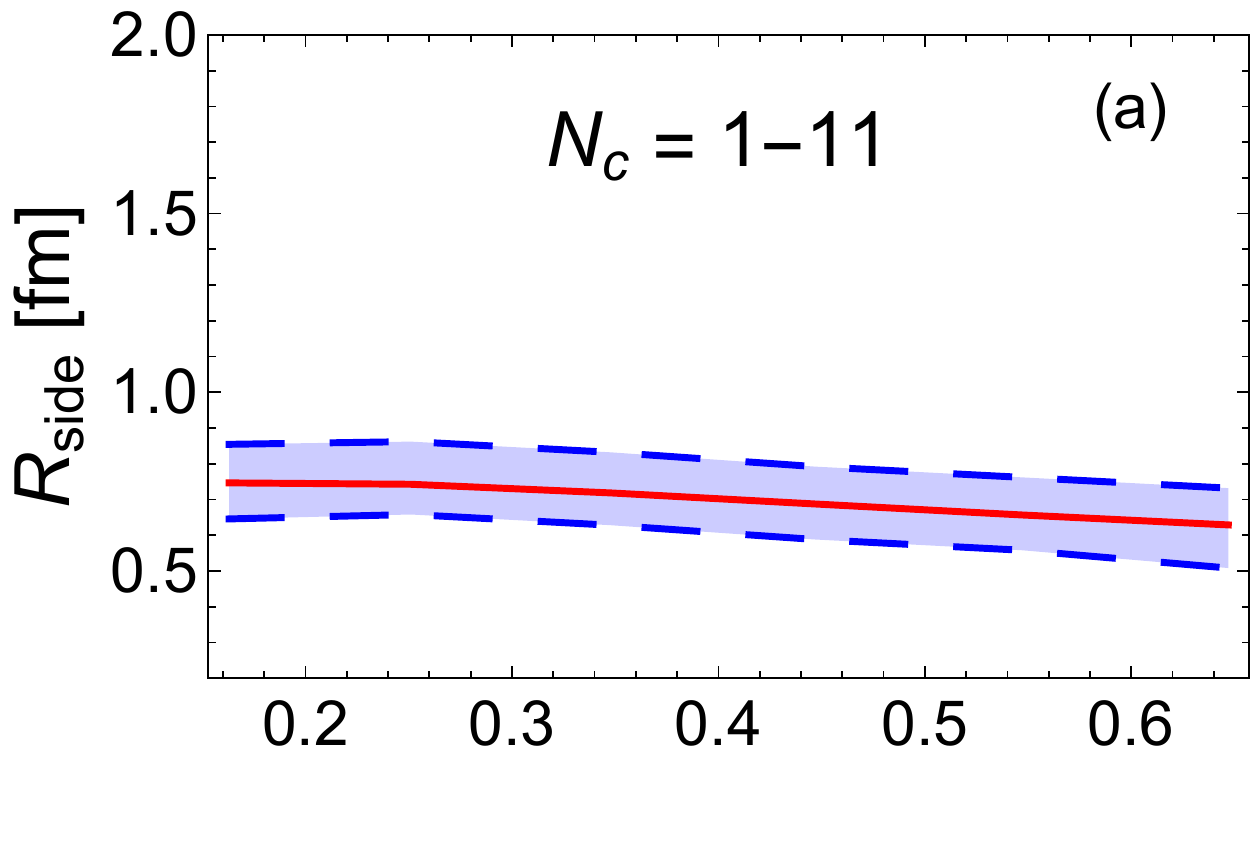}  \quad
 \includegraphics[scale=0.33]{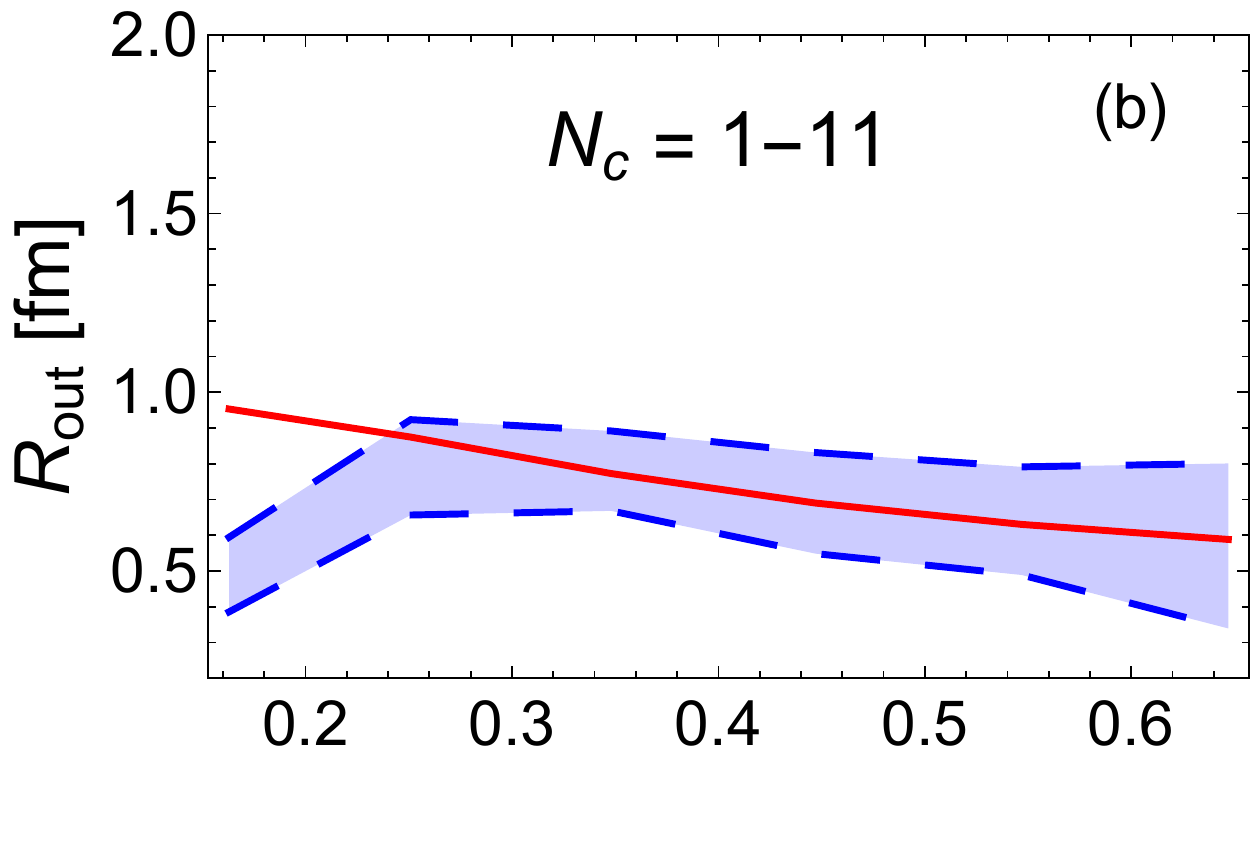} \quad
\includegraphics[scale=0.33]{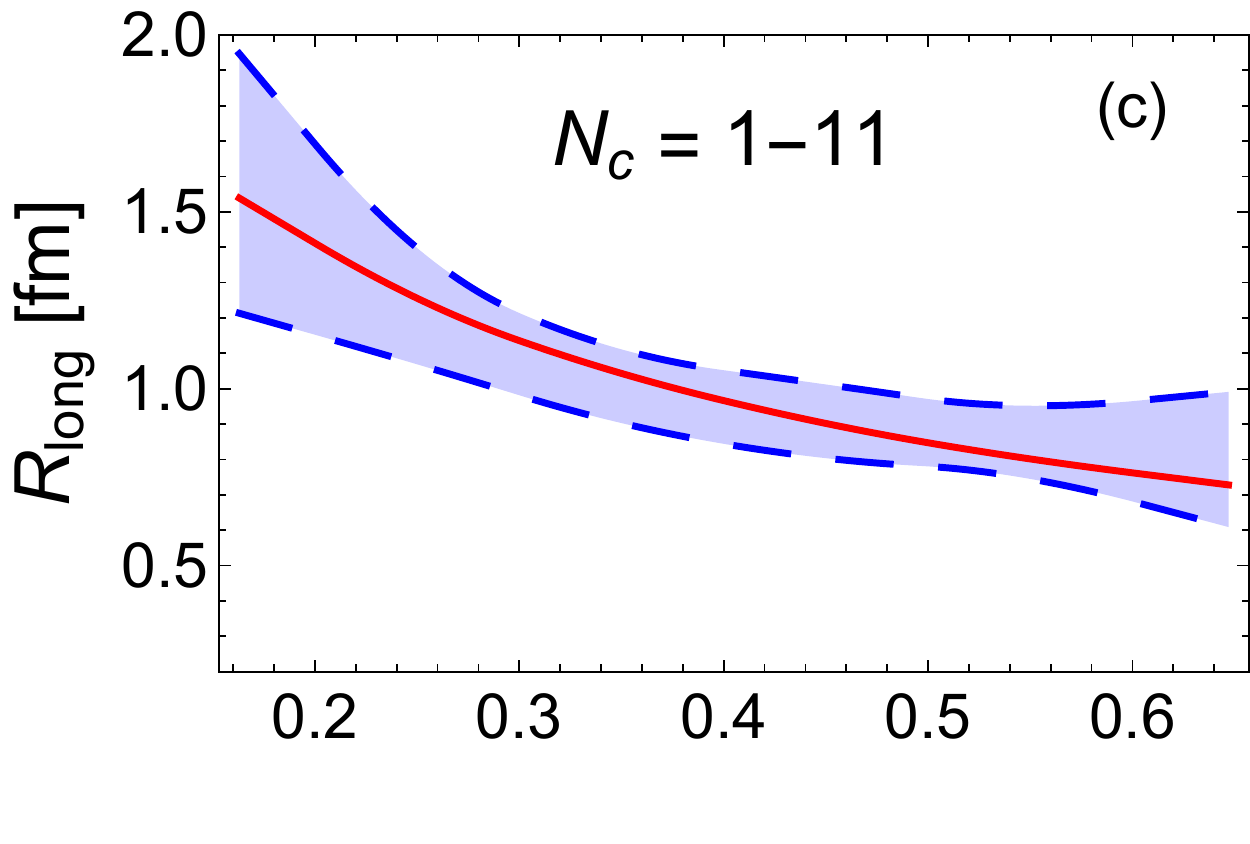}  \\
\includegraphics[scale=0.33]{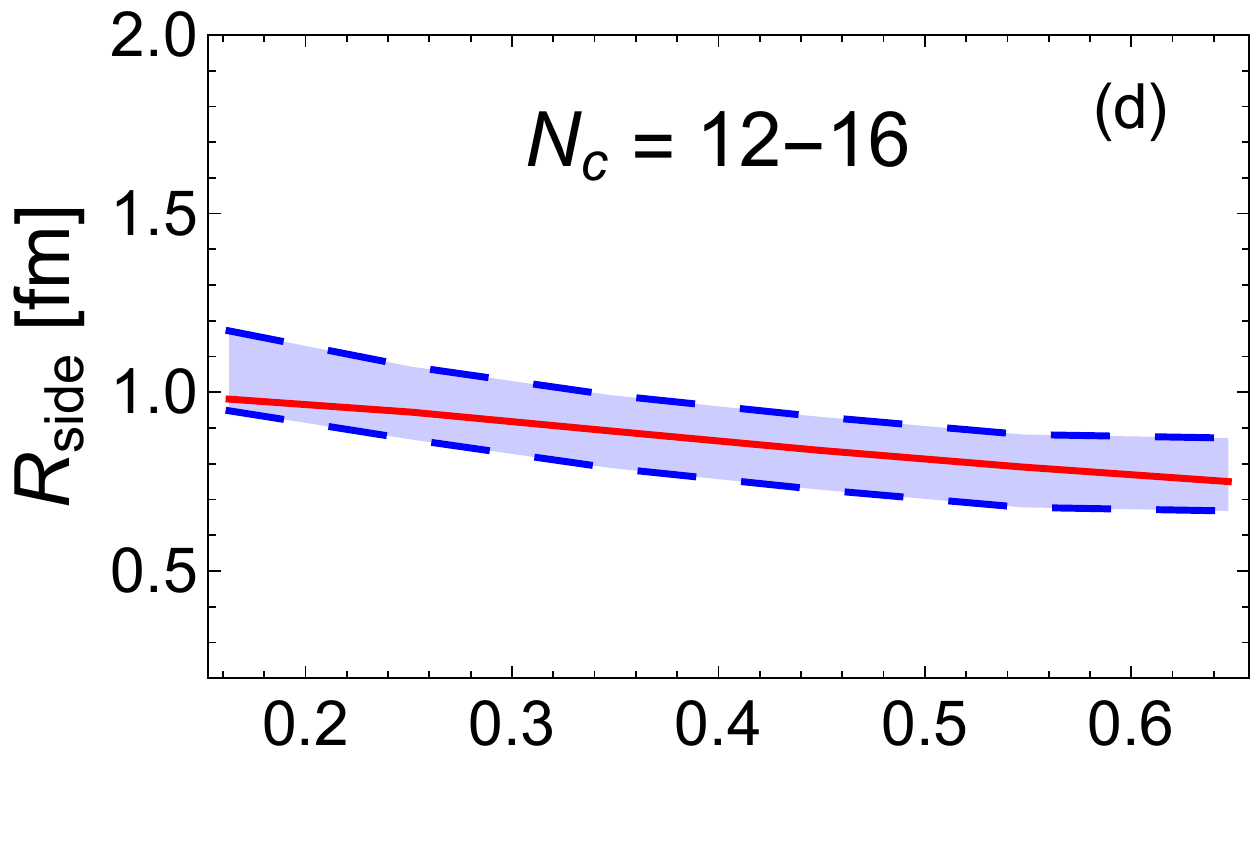}  \quad
 \includegraphics[scale=0.33]{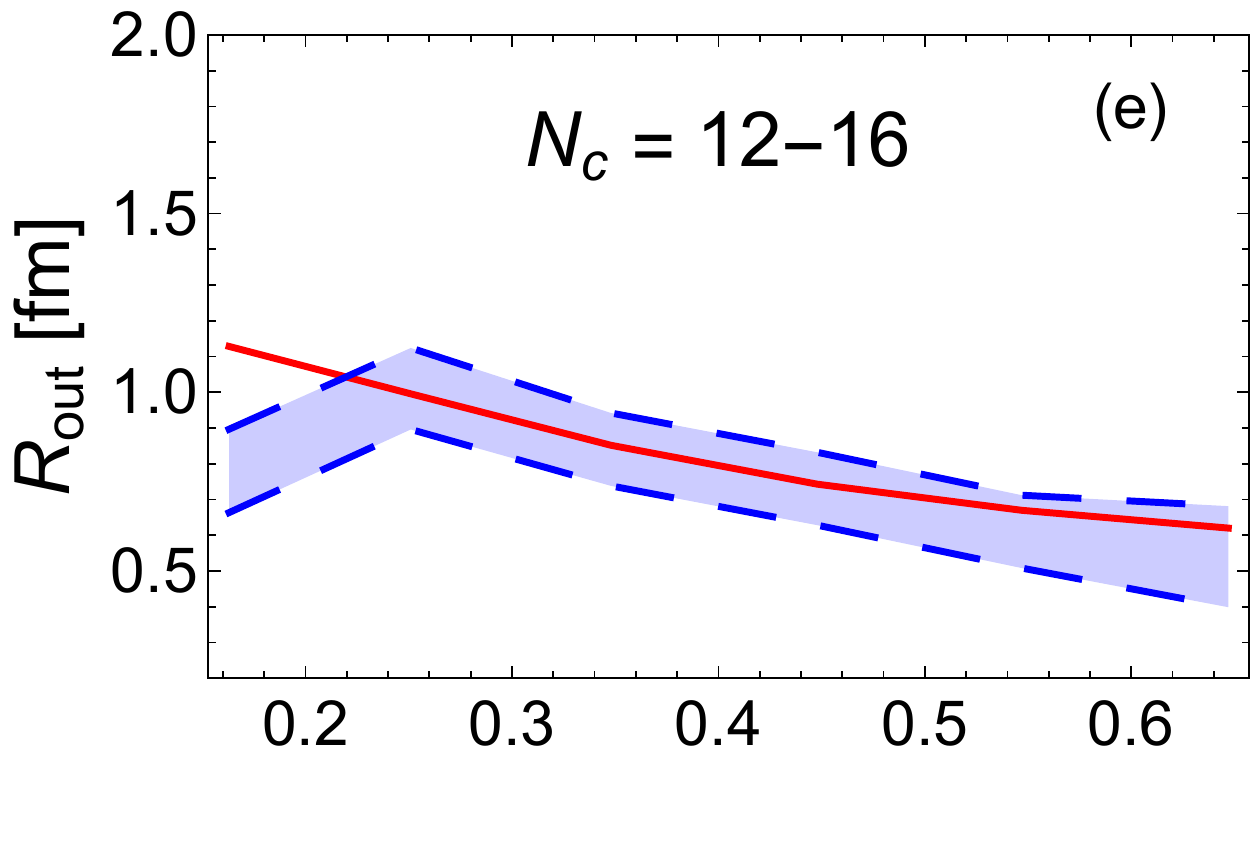} \quad
\includegraphics[scale=0.33]{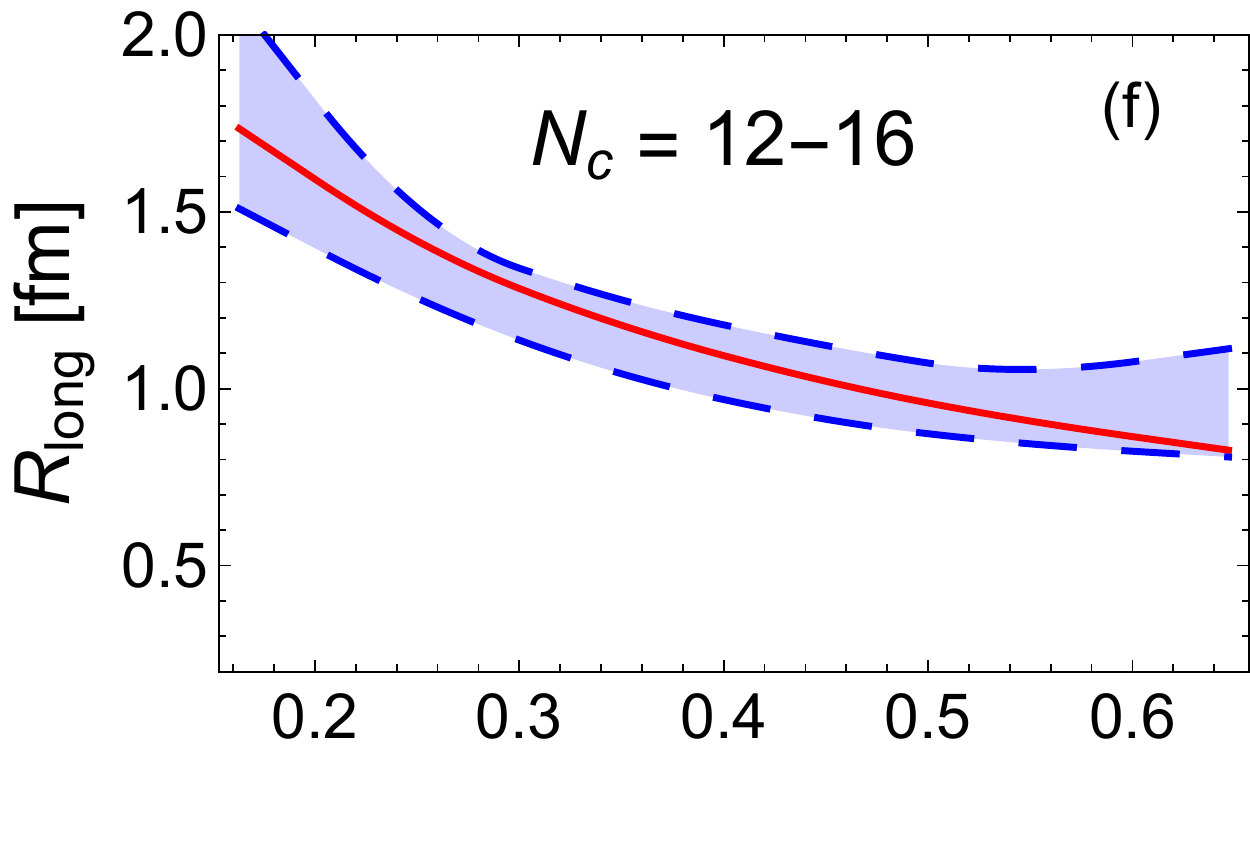}  \\
\includegraphics[scale=0.33]{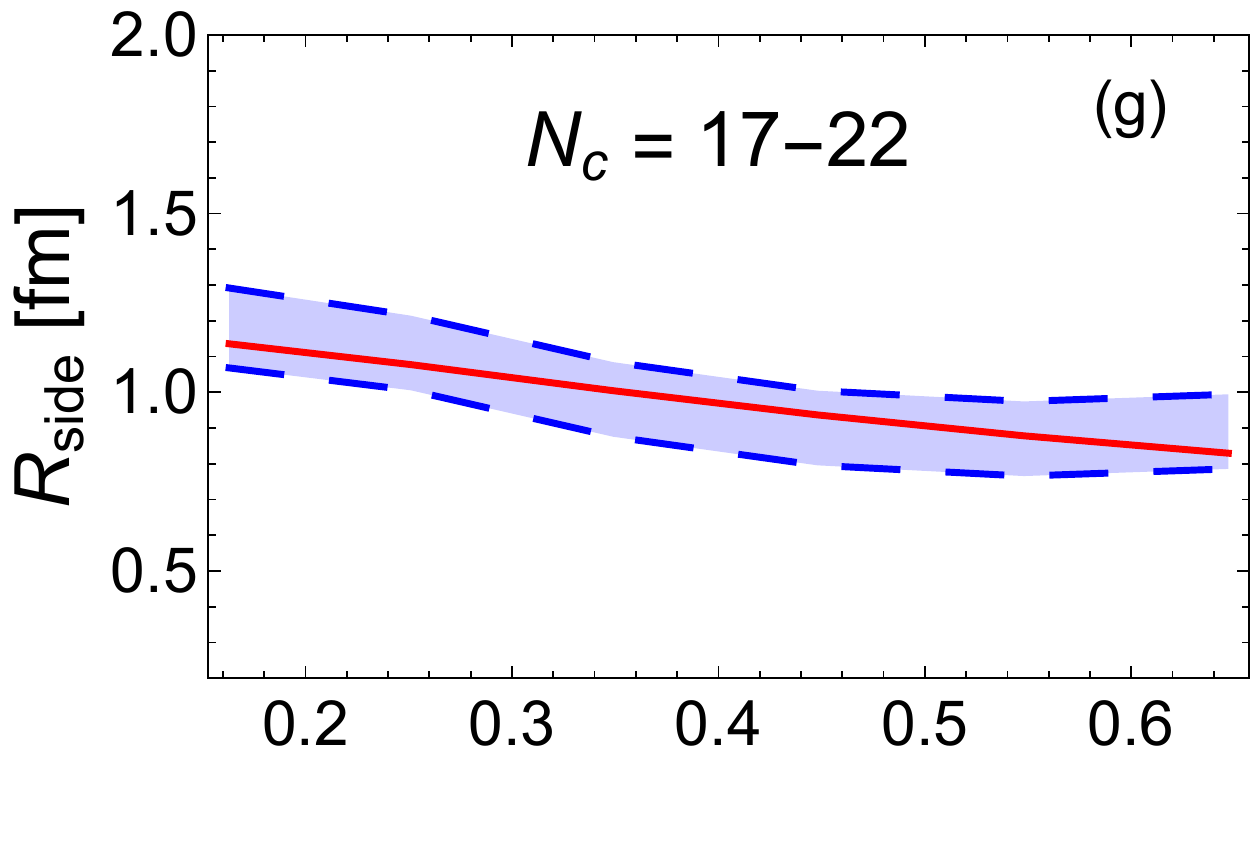}  \quad
 \includegraphics[scale=0.33]{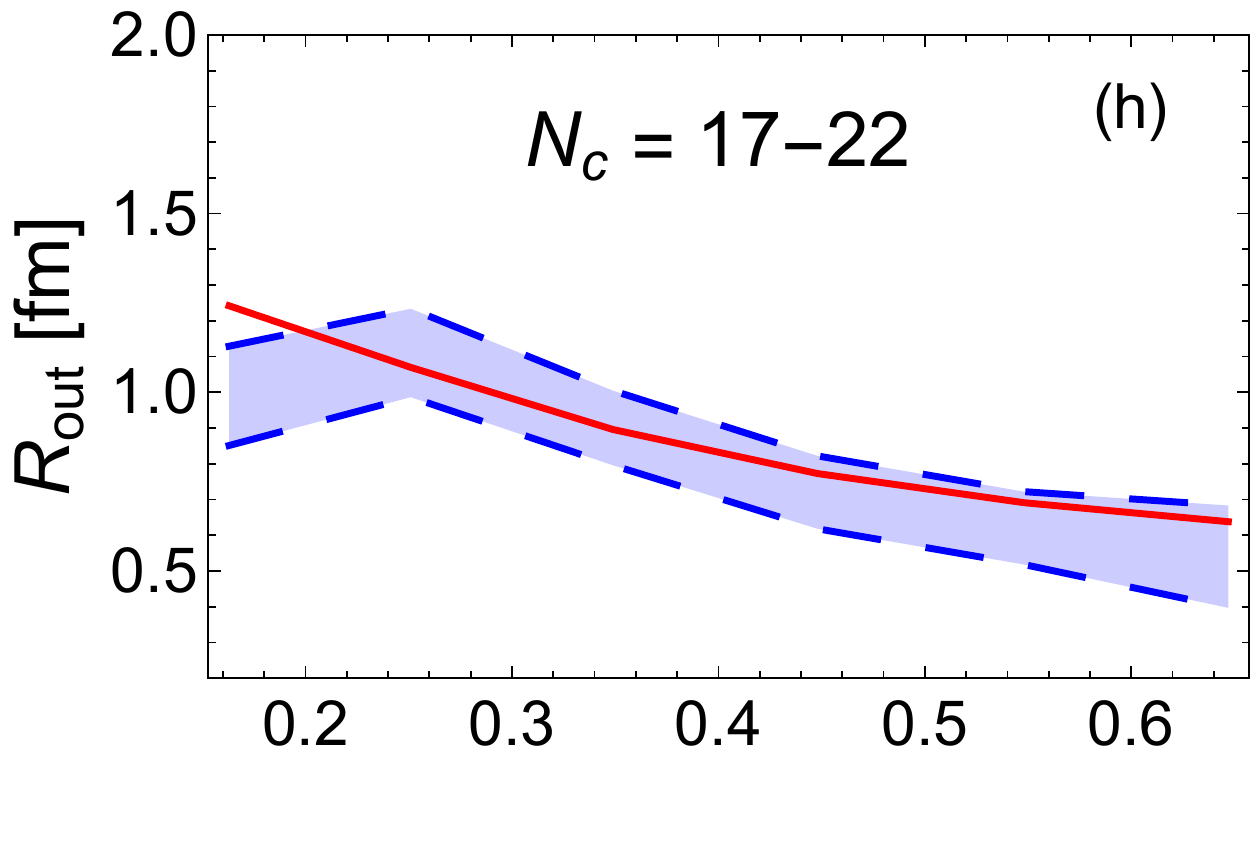} \quad
\includegraphics[scale=0.33]{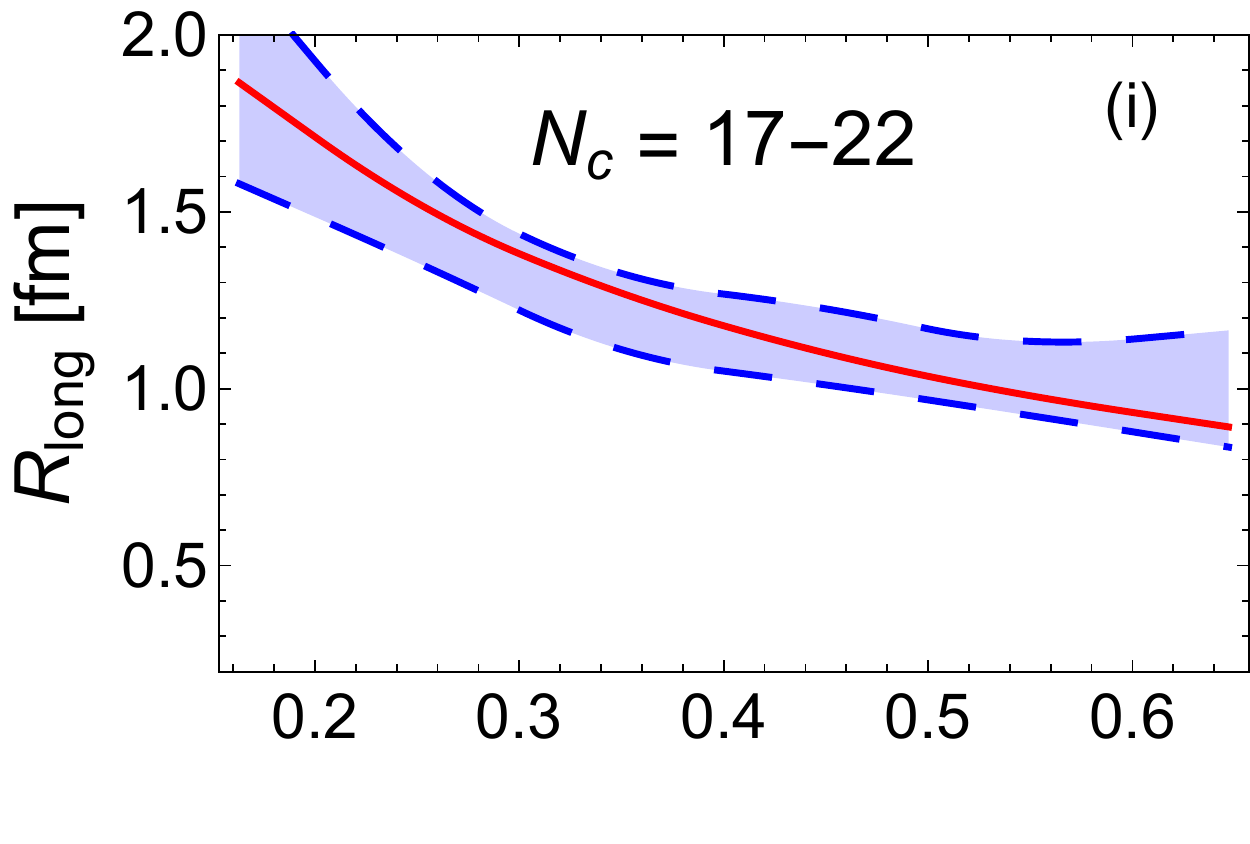}  \\
\includegraphics[scale=0.33]{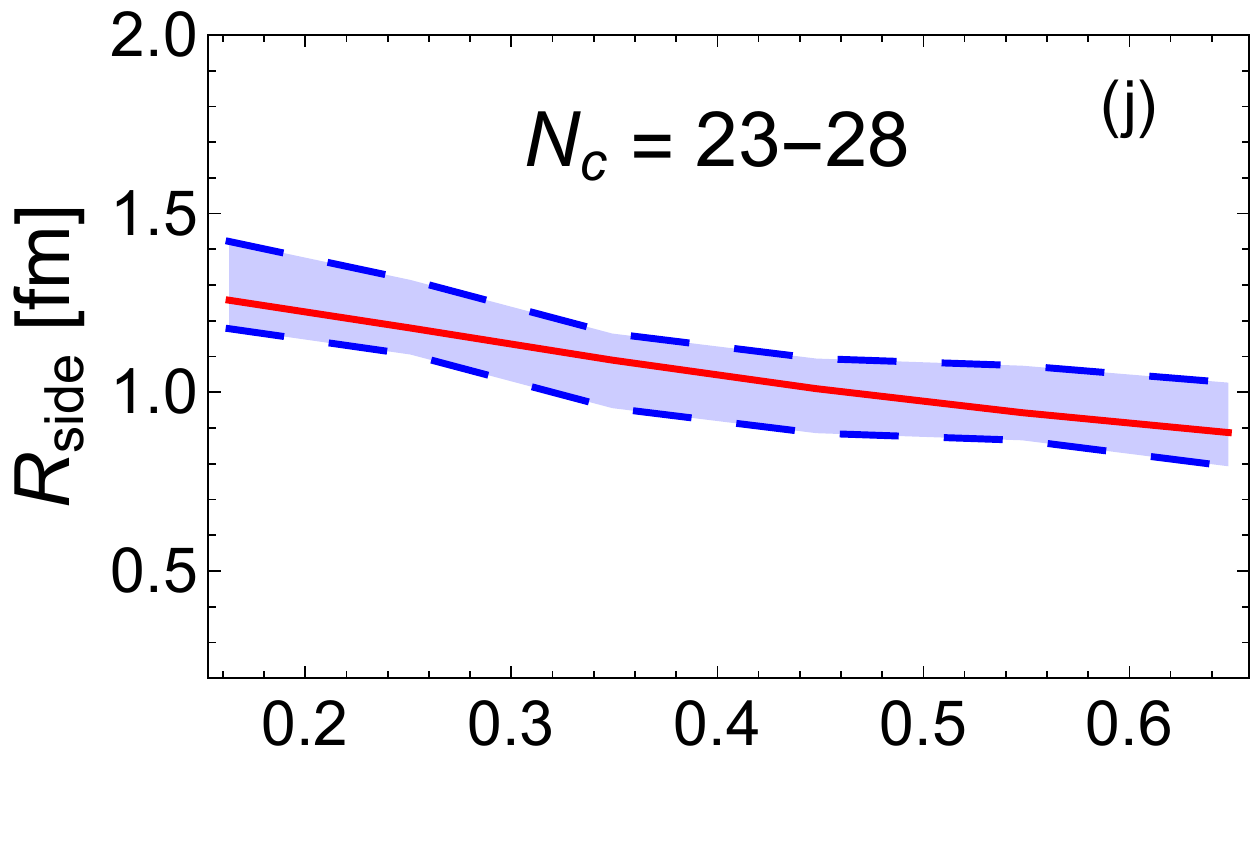}  \quad
 \includegraphics[scale=0.33]{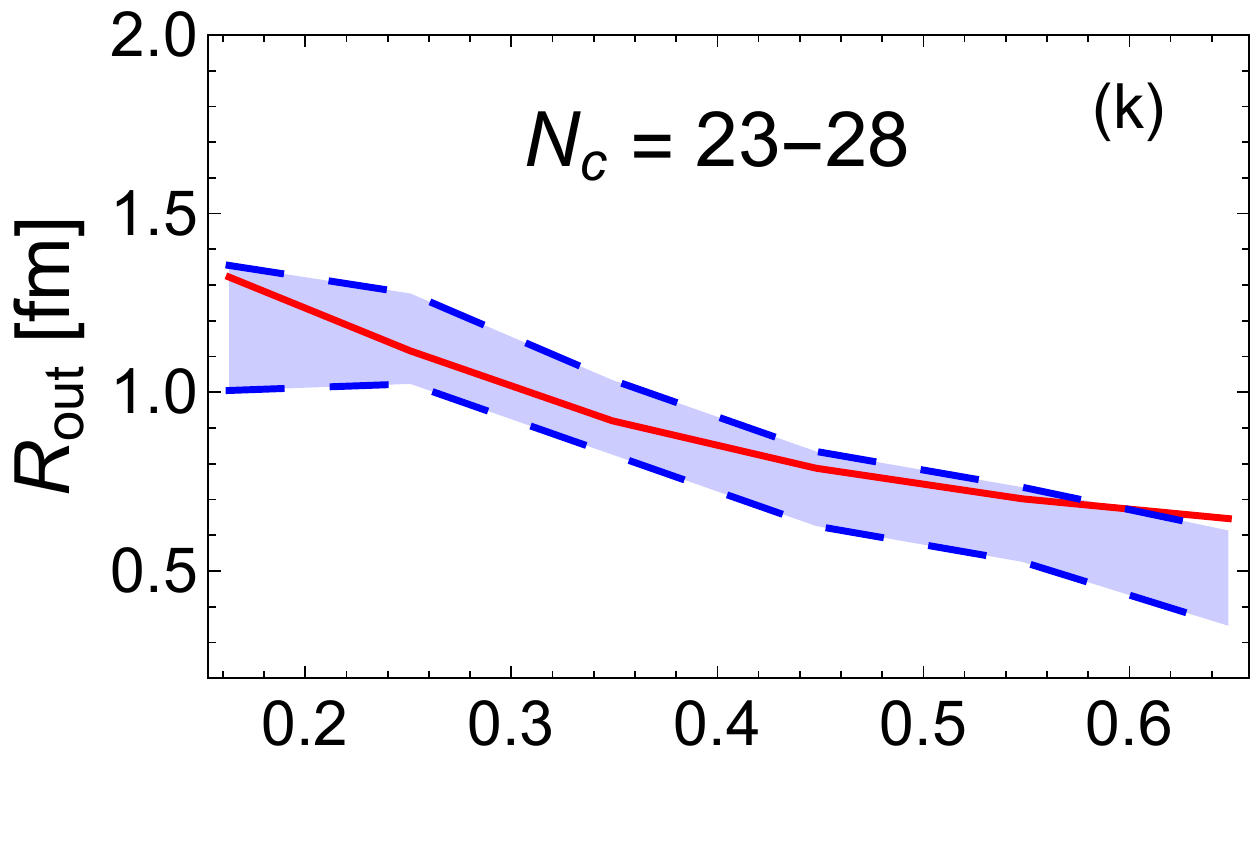} \quad
\includegraphics[scale=0.33]{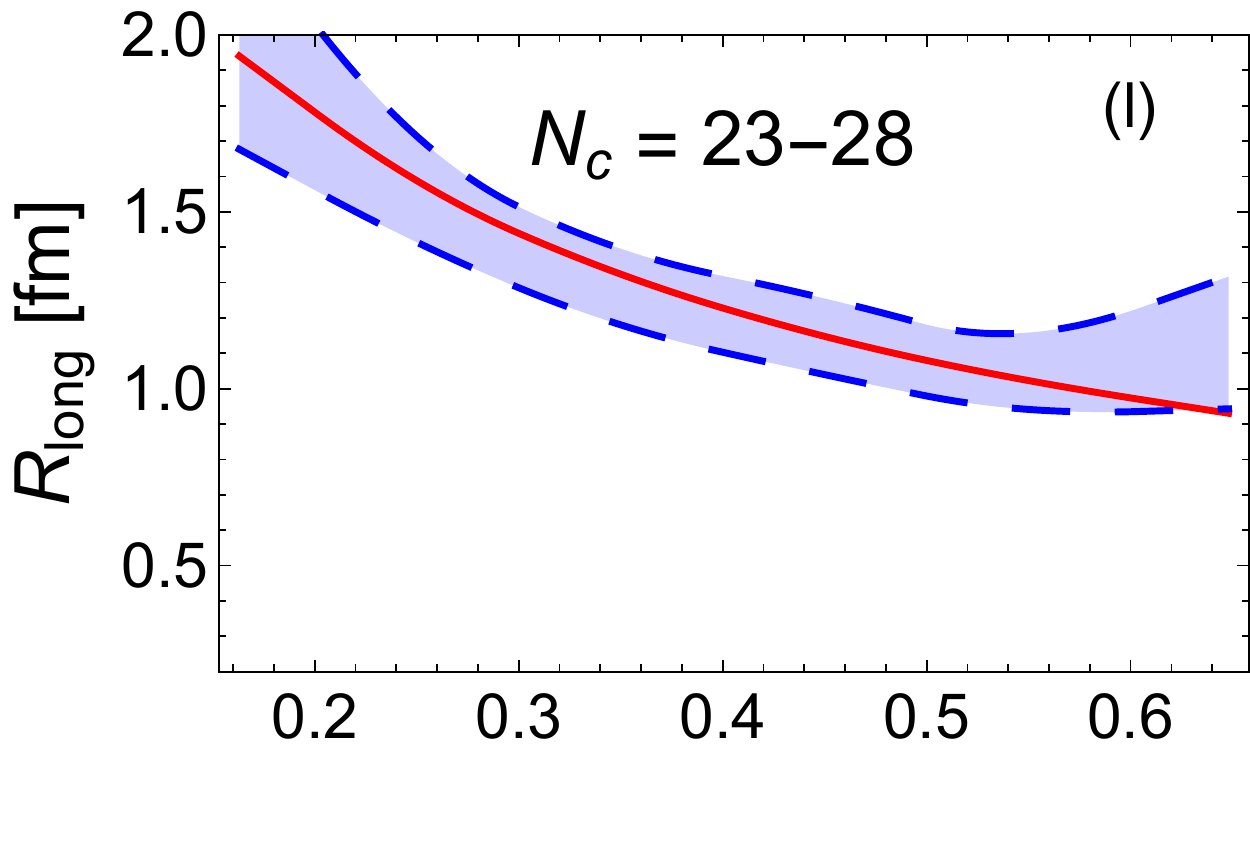}  \\
\includegraphics[scale=0.33]{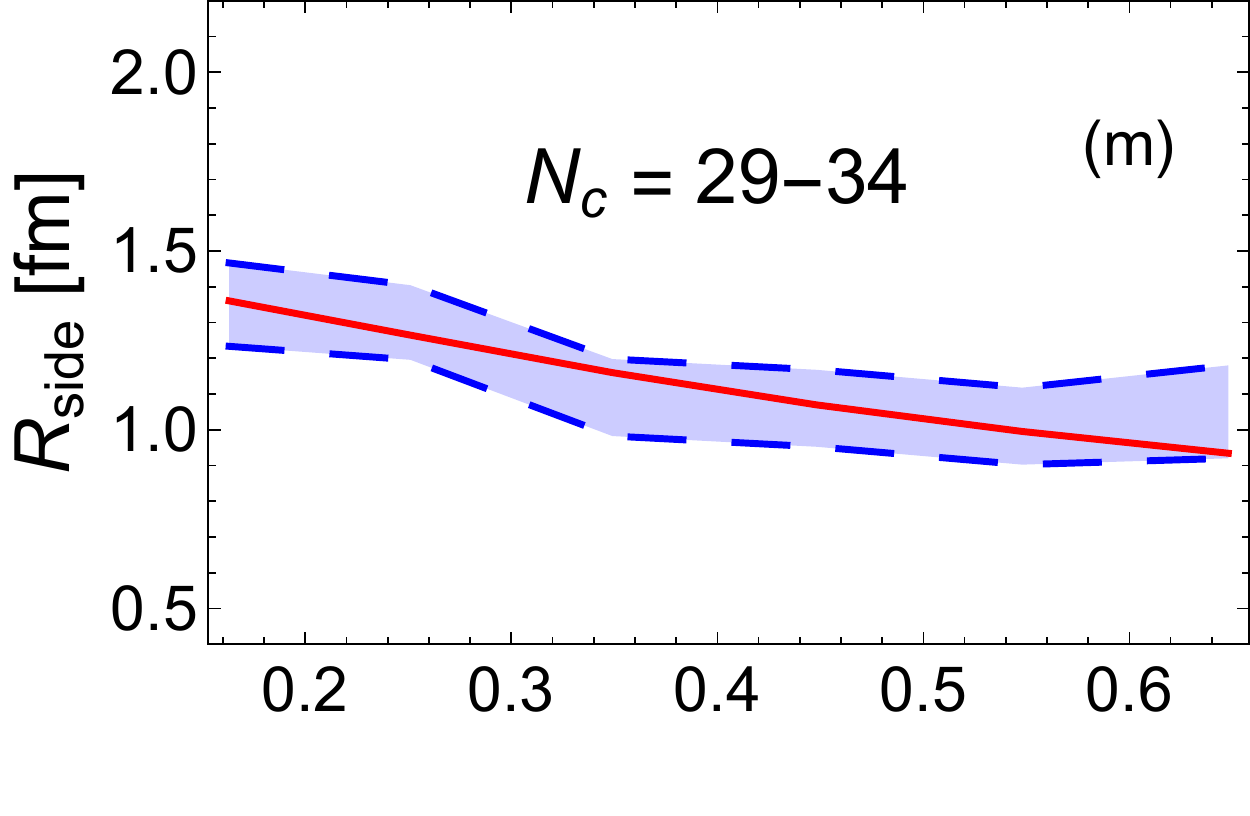}  \quad
 \includegraphics[scale=0.33]{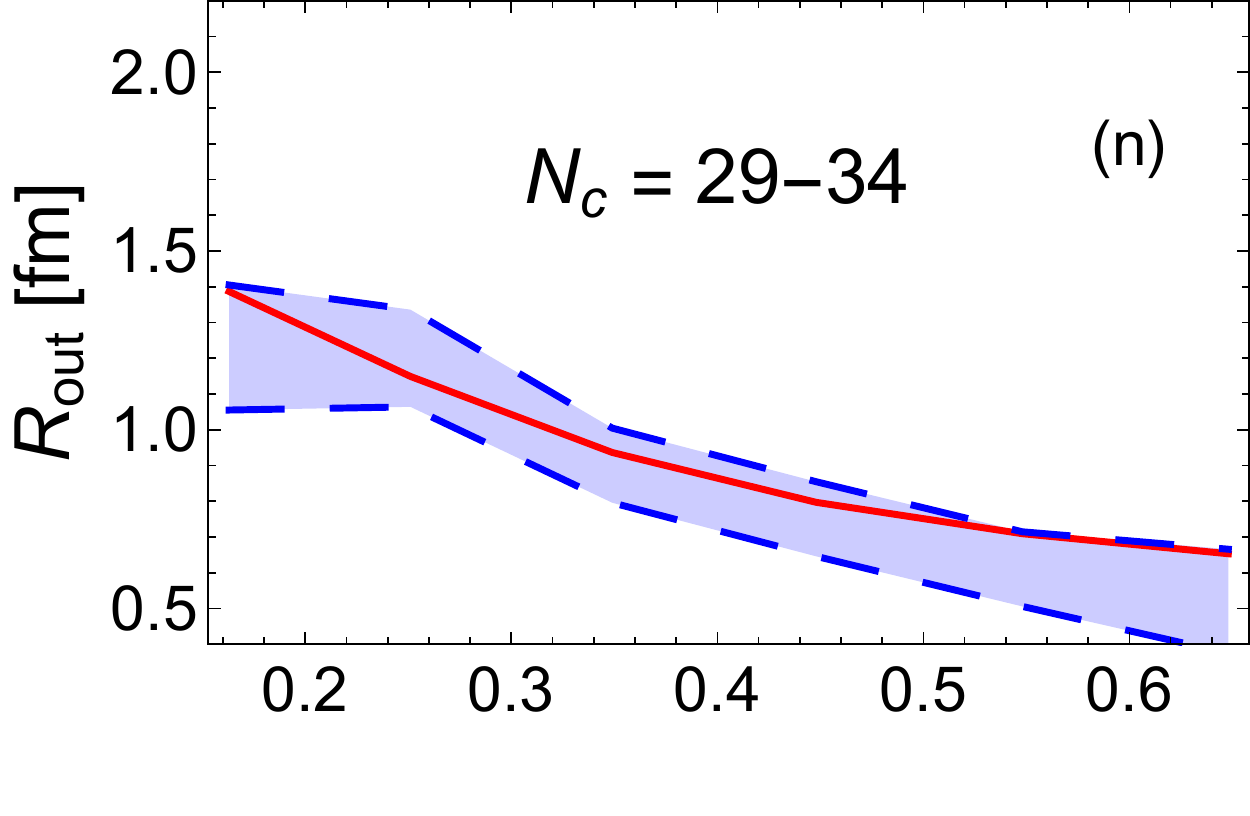} \quad
\includegraphics[scale=0.33]{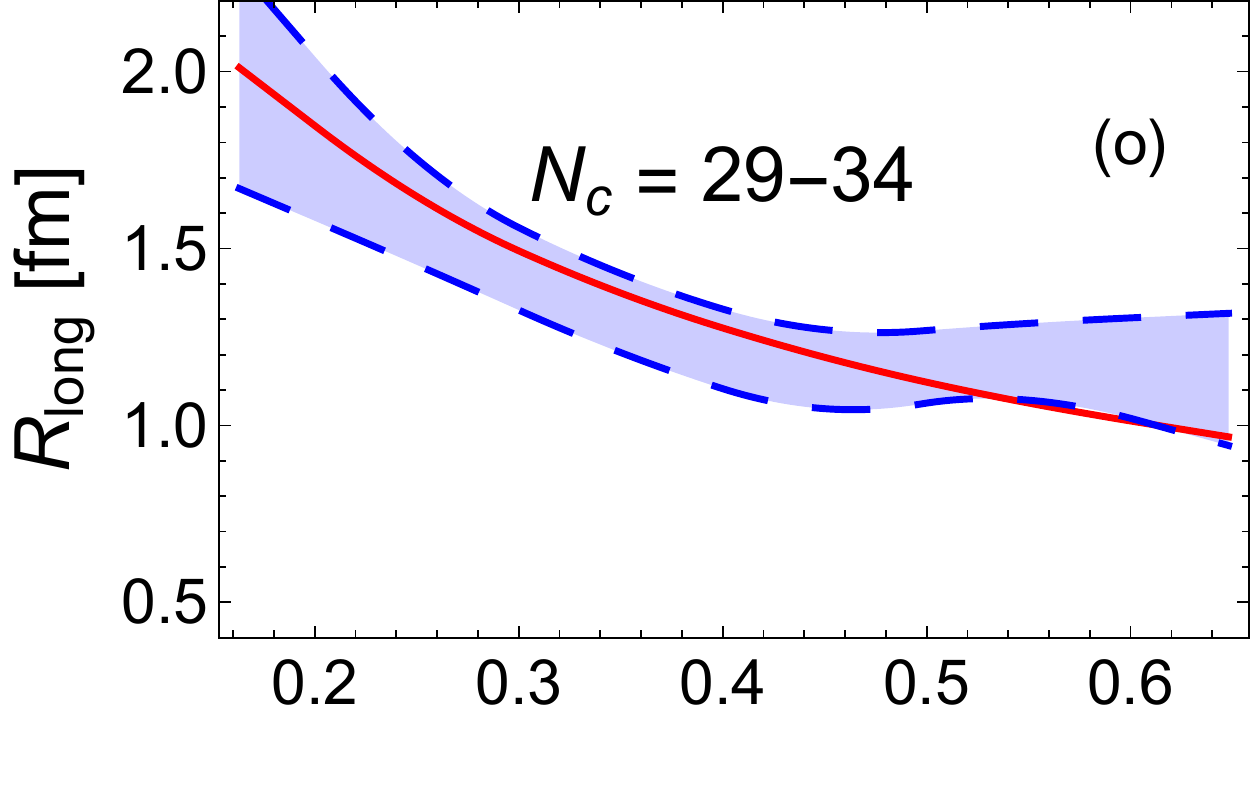}  \\
\includegraphics[scale=0.33]{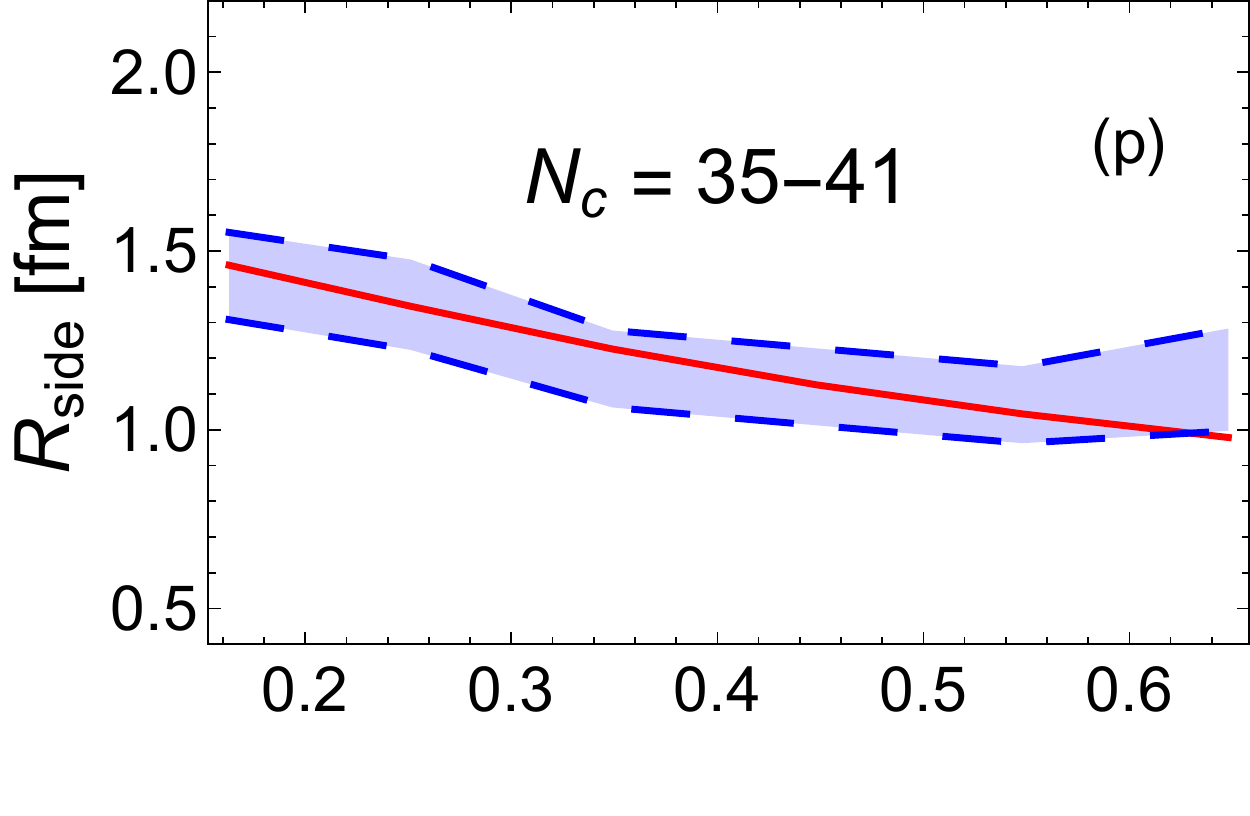}  \quad
 \includegraphics[scale=0.33]{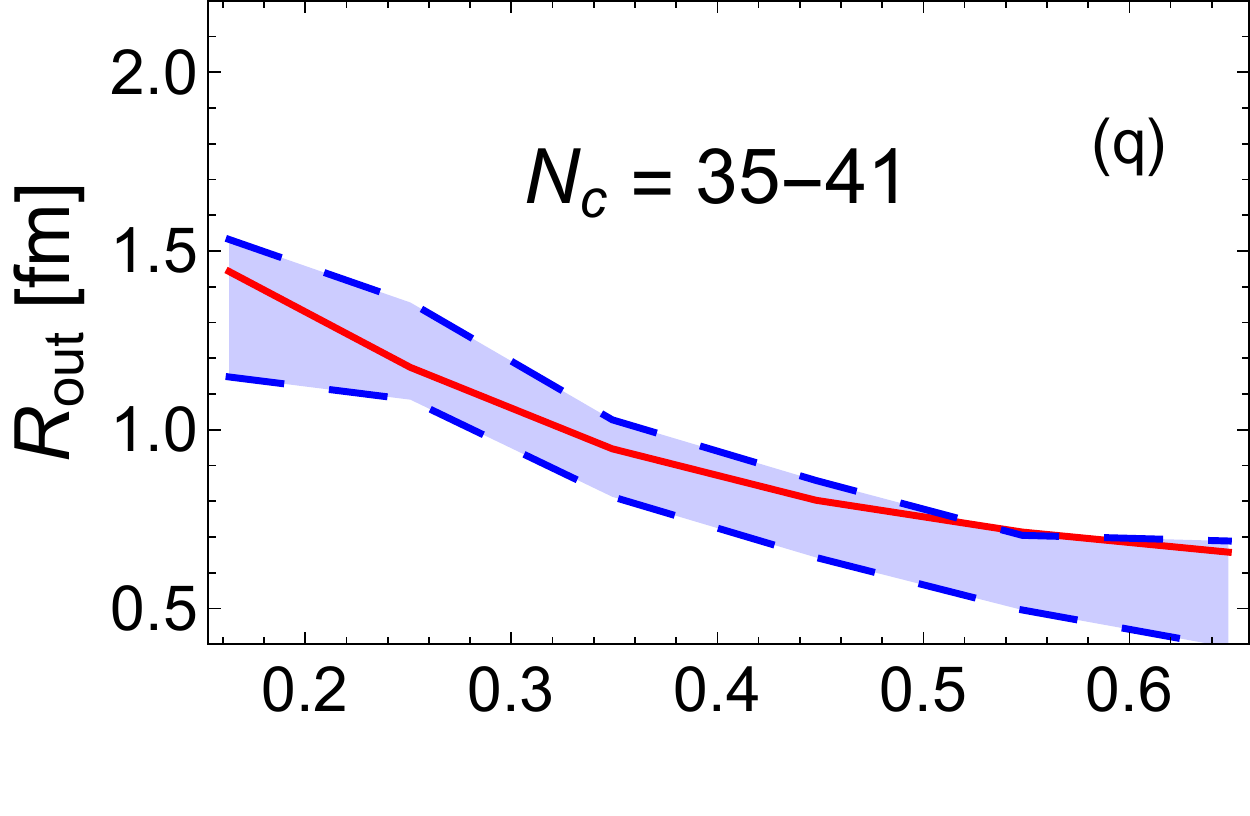} \quad
\includegraphics[scale=0.33]{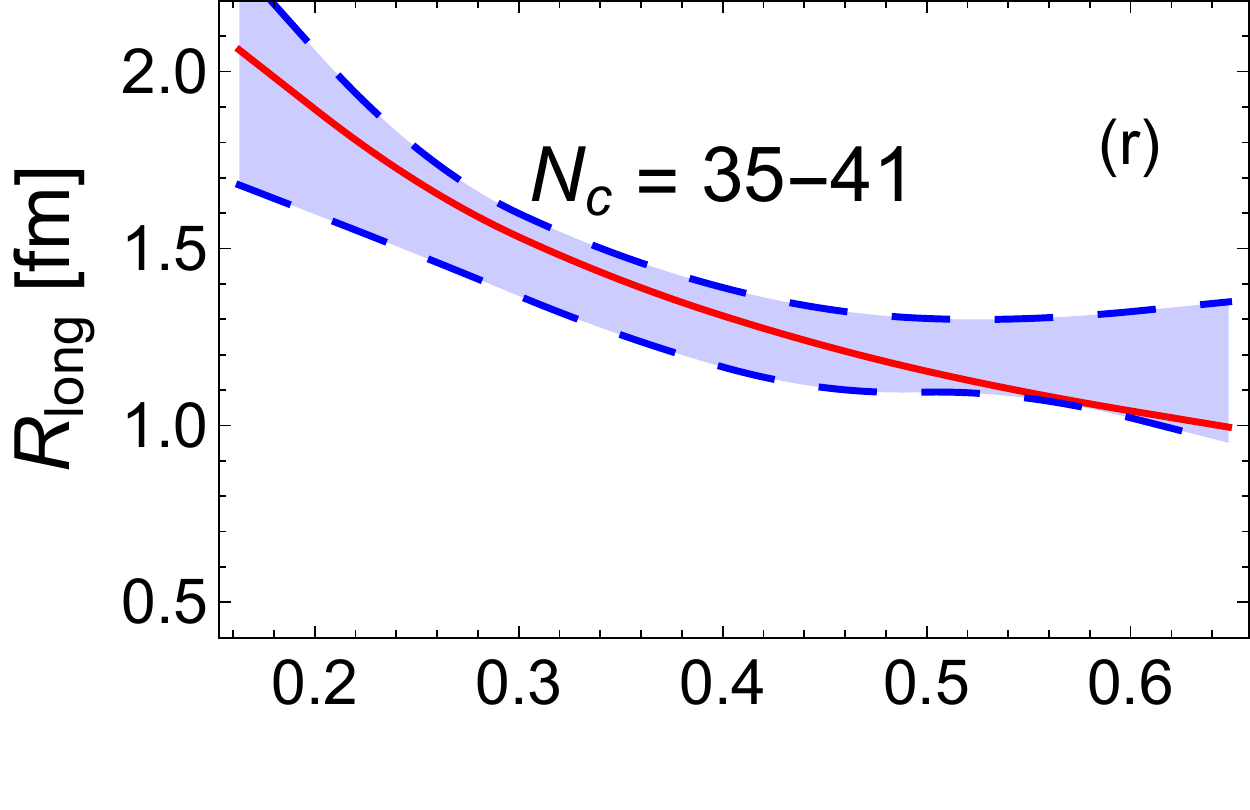}  \\
\includegraphics[scale=0.33]{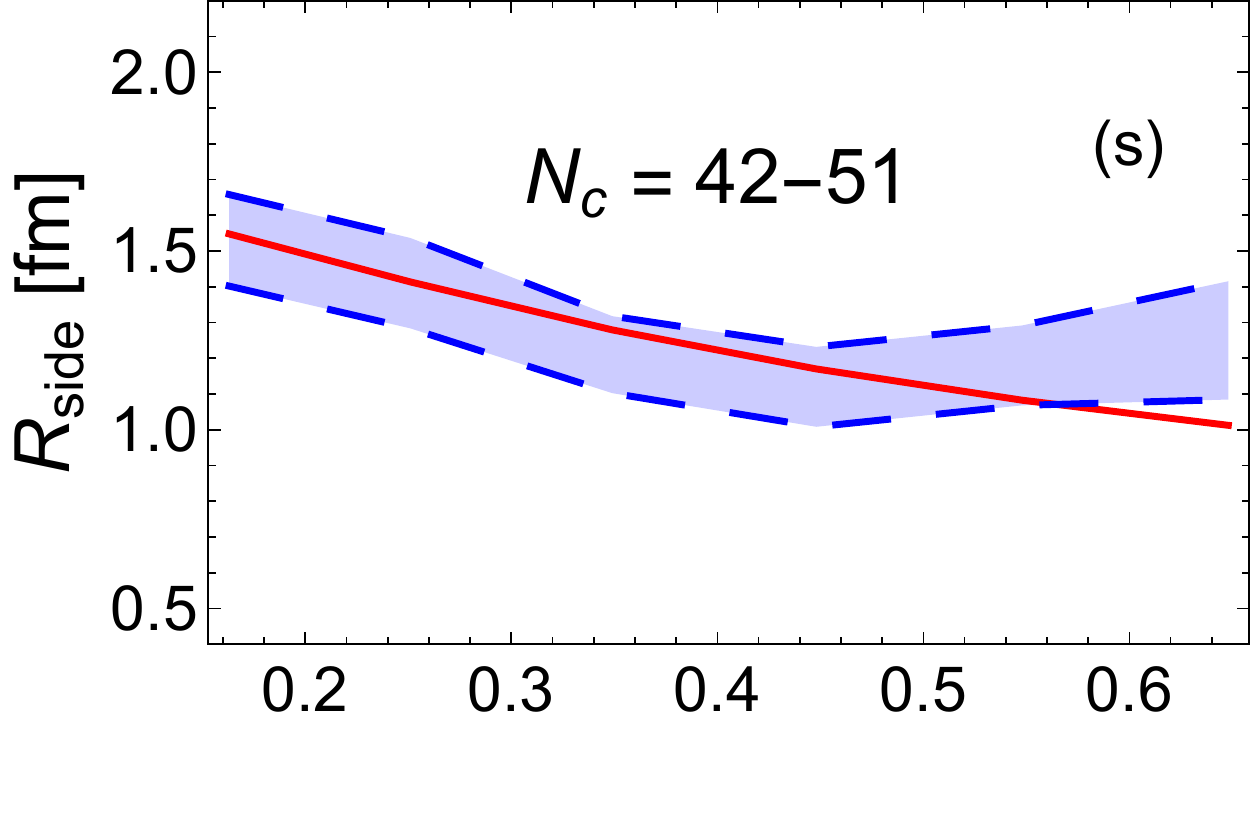}  \quad
 \includegraphics[scale=0.33]{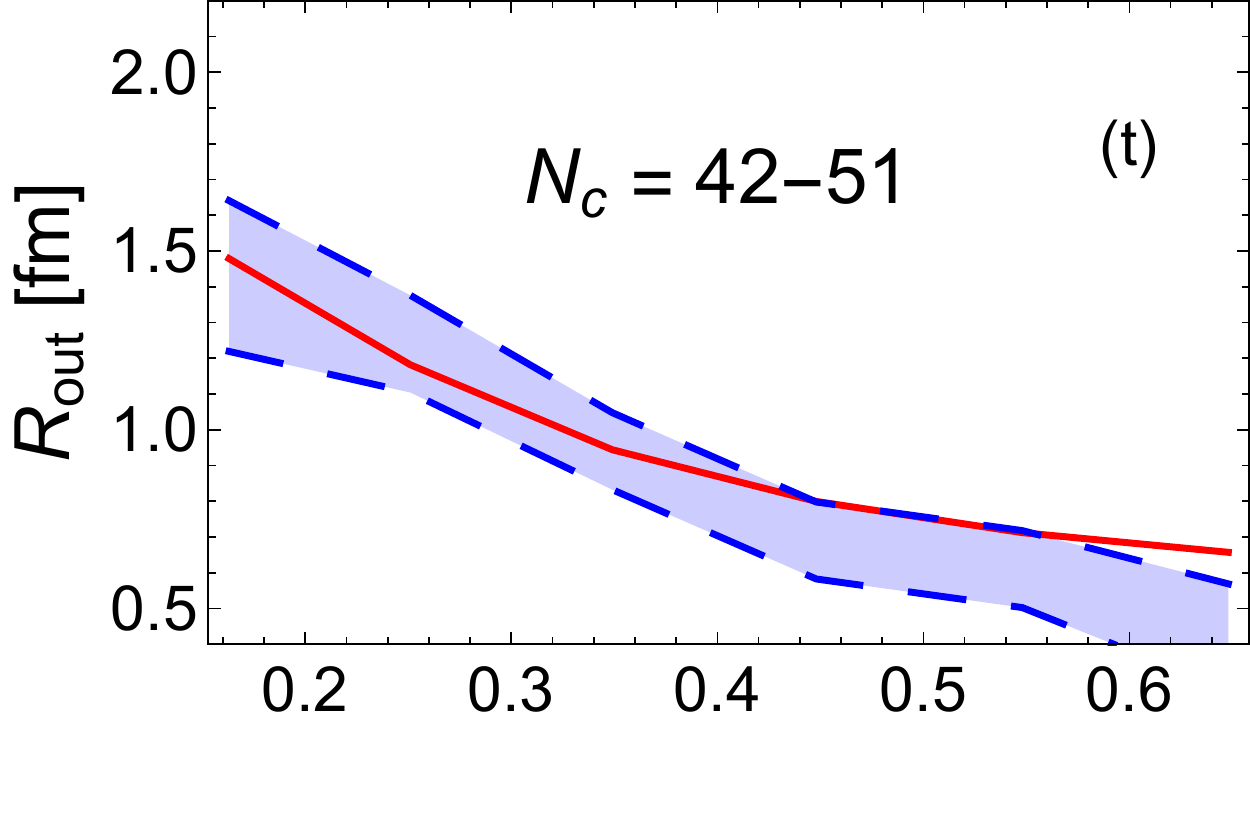} \quad
\includegraphics[scale=0.33]{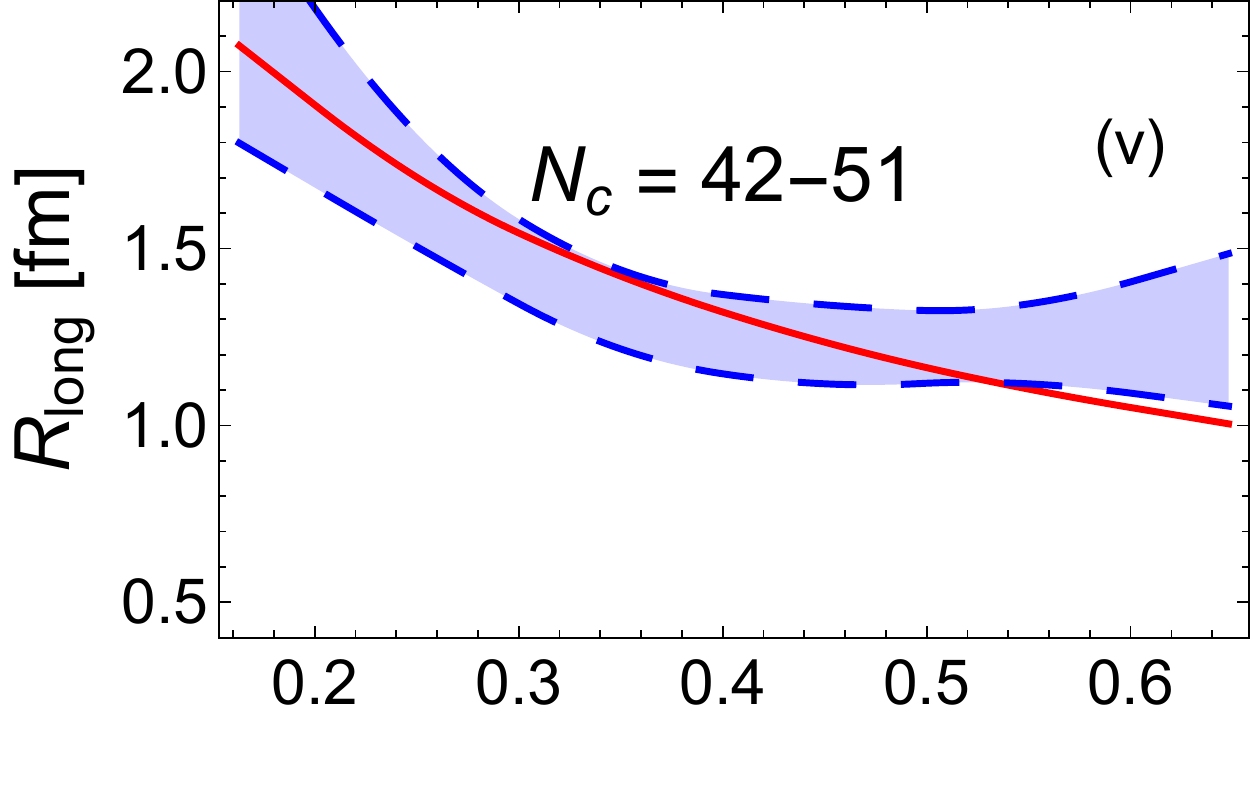}  \\
\includegraphics[scale=0.33]{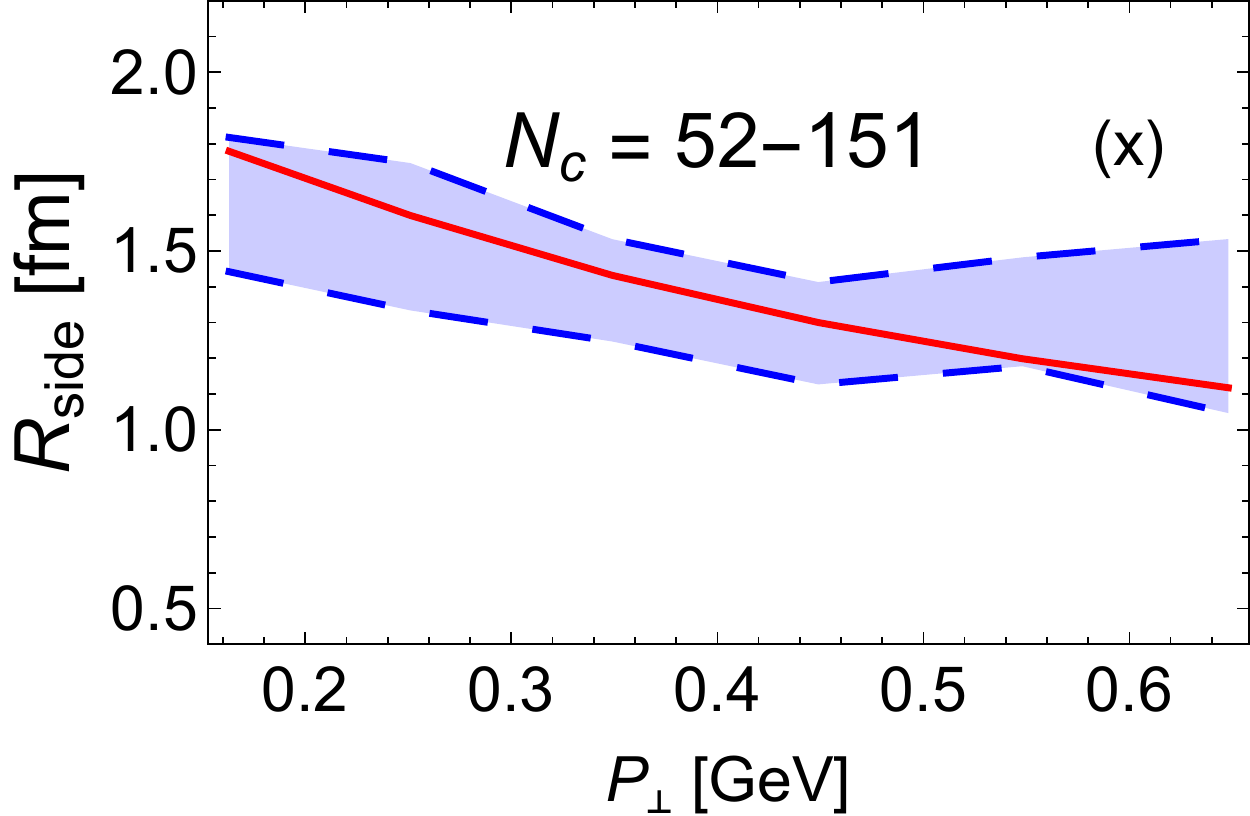}  \quad
 \includegraphics[scale=0.33]{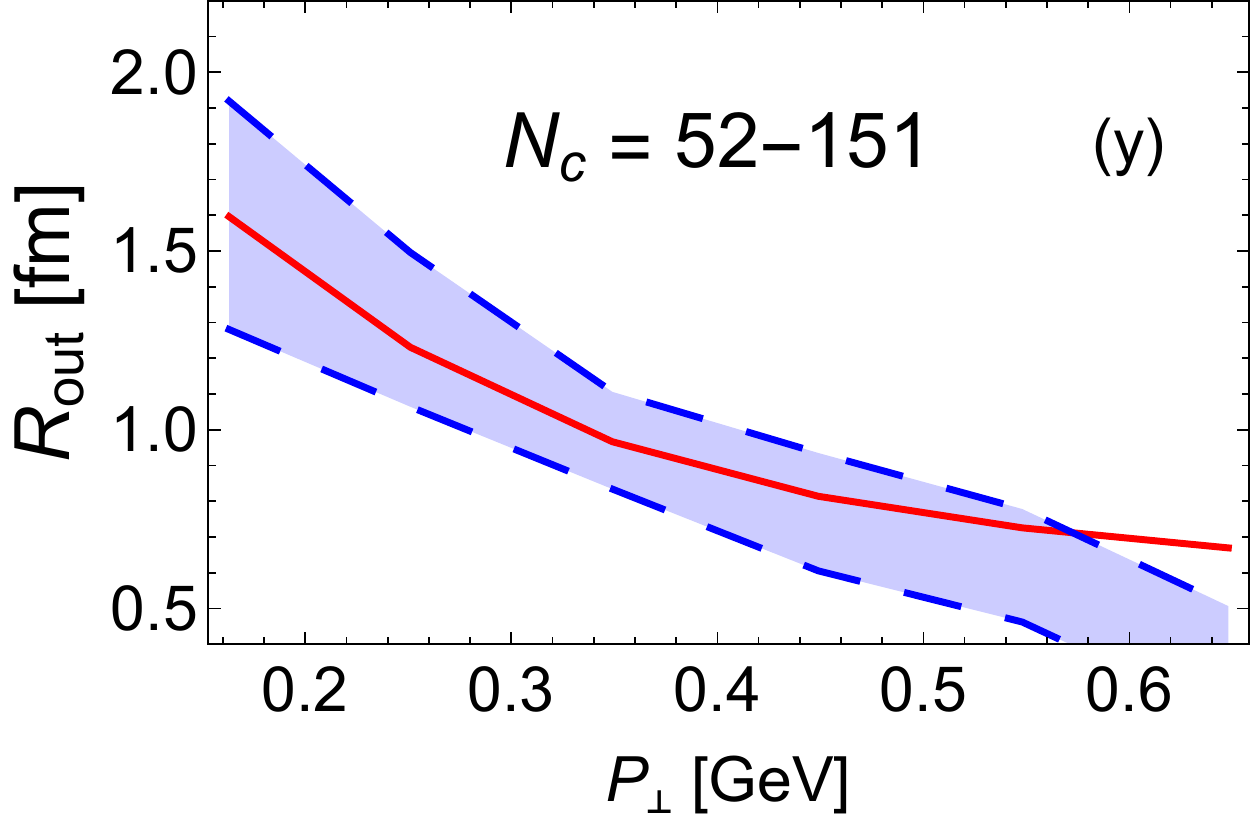} \quad
\includegraphics[scale=0.33]{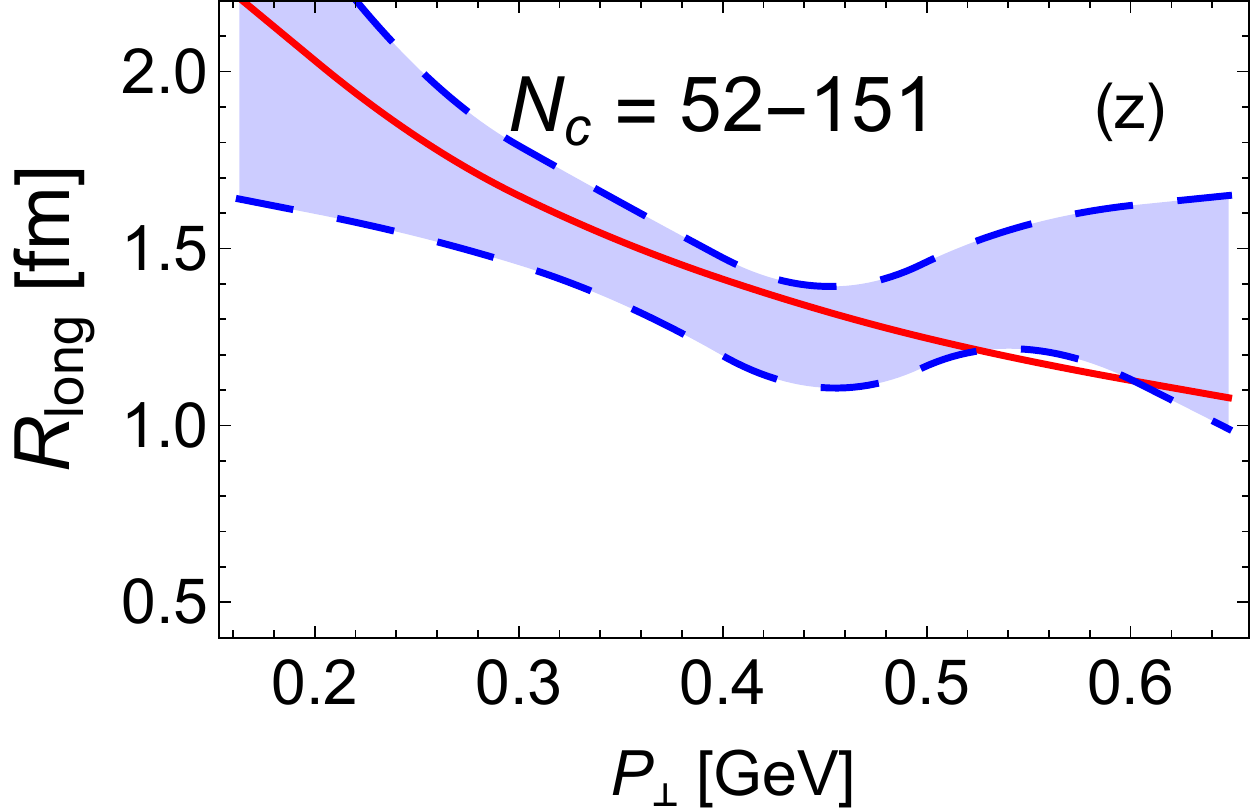}  \\\end{center}
\caption{(Color online) The model results (red solid curves) compared to the experiment results (central points of the blue bands).  The width of the bands represents the experimental error. }%
\label{radii}%
\end{figure}

\begin{figure}[b]
\begin{center}
\includegraphics[scale=0.5]{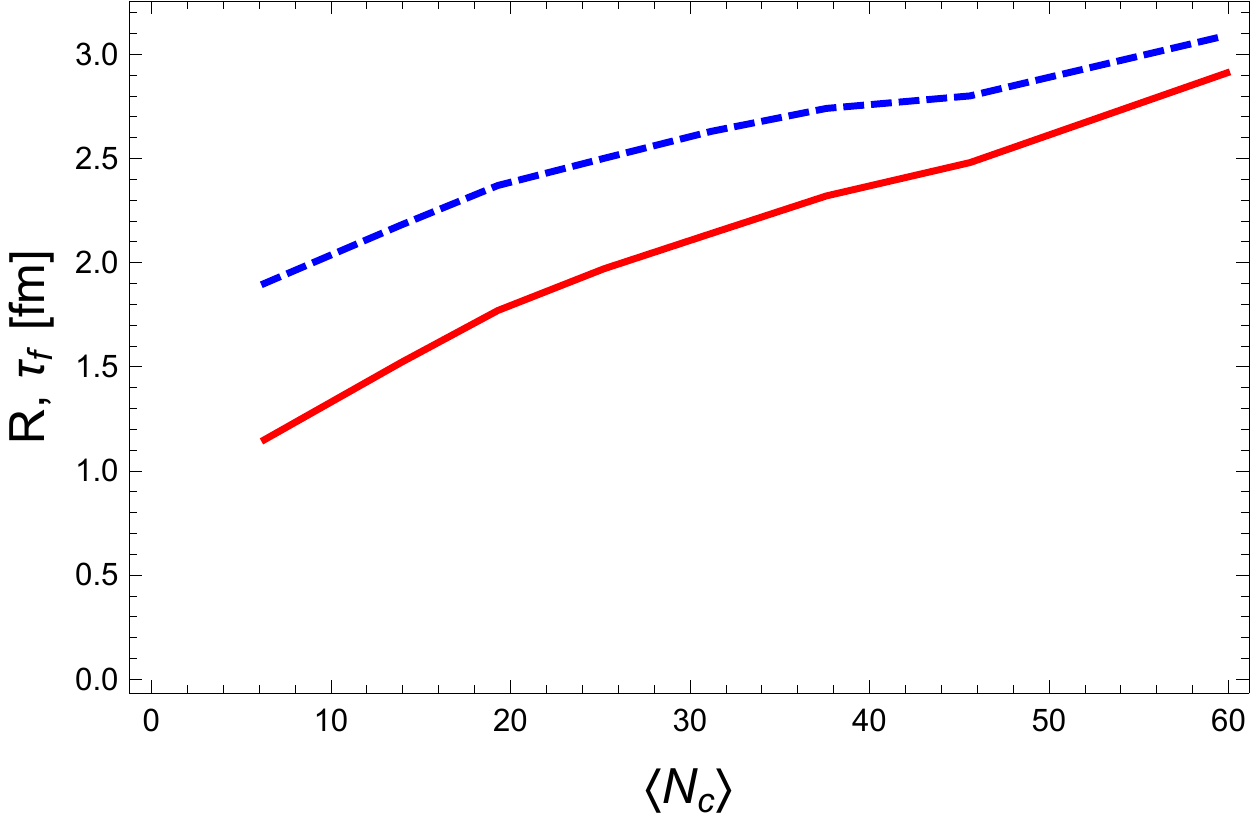} 
\end{center}
\caption{(Color online) The fitted values of $R$ (solid line) and $\tau_f$ (dashed line) as functions of the mean multiplicity, see Table~\ref{par}.}%
\label{pt23}%
\end{figure}

To determine $R$ and $\t$ we minimized $\chi^2$, using  5 intervals of $P_\p$ (the lowest one was omitted for  reasons explained below). The results are summarized in  Fig.~\ref{radii} where one sees  that this procedure gives a rather  good description of data. With the value of $\d$ fixed at 0.75 fm, the results for $R$ and $\t$ are presented in Table~\ref{par}.
\begin{table}[b]
{\begin{tabular}{ccccccccc} 
\hline 
mult. class & 1--11 &  12--16  &  17--22 &  23--28 & 29--34 &  35--41 & 42--51 & 52--151 \\ 
\hline \\
$\langle N_c \rangle$        &    6.3  &     13.9  &     19.3  &    25.2  &   31.2  &     37.6  &    45.6 &  59.9 \\
$R$  [fm]          &  1.15  &    1.52   &    1.77   &    1.97  &   2.14  &     2.32  &    2.49 &  2.91 \\
$\t$ [fm]     &  1.90 &   2.18 &  2.37 & 2.50 &  2.63 &  2.74 &  2.80 &  3.09 \\
$\chi^2$    &  0.96&  1.90&  2.89&   4.06&  5.88&   5.45& 11.63&  8.48 \\
$\chi^2_{\rm tot}$ &18.77&   11.08&   6.34&  4.86&   6.65&  5.79&  11.72&  9.7
\end{tabular}
}
\caption{Model parameters and $\chi^2$ values for different multiplicity classes.}
\label{par}
\end{table}
The multiplicity dependence of the parameters $R$ and $\tau_f$ is also shown in Fig.~\ref{pt23}.  In the fifth row of  Table~\ref{par} we show the values of $\chi^2$ (not divided by the number of degrees of freedom which is 13 in this case). The values of $\chi^2$ indicate that the deviations from the experimental values are indeed very small. With increasing $\d$ the description becomes worse, but it is still acceptable up to $\d=0.85$~fm (these results are not presented here).

When the smallest $P_\p$ bin is included, the description is worse. The corresponding values of $\chi^2$, denoted $\chi^2_{\rm tot}$,  are shown in the sixth row of Table~\ref{par}. One sees from  Fig.~\ref{radii} that  the discrepancy is due to the bad description of $R_{\rm out}$ at smallest $P_\p$. Indeed, the data show an anomalous behavior: at small multiplicities $R_{\rm out}$ {\it increases with $P_\p$}, whereas the model predicts a steady decrease. A possible explanation of this "anomalous" effect is discussed in Sec.~\ref{sect:corneq0}. 

\FloatBarrier

\begin{figure}[b]
\begin{center}
\includegraphics[scale=0.5]{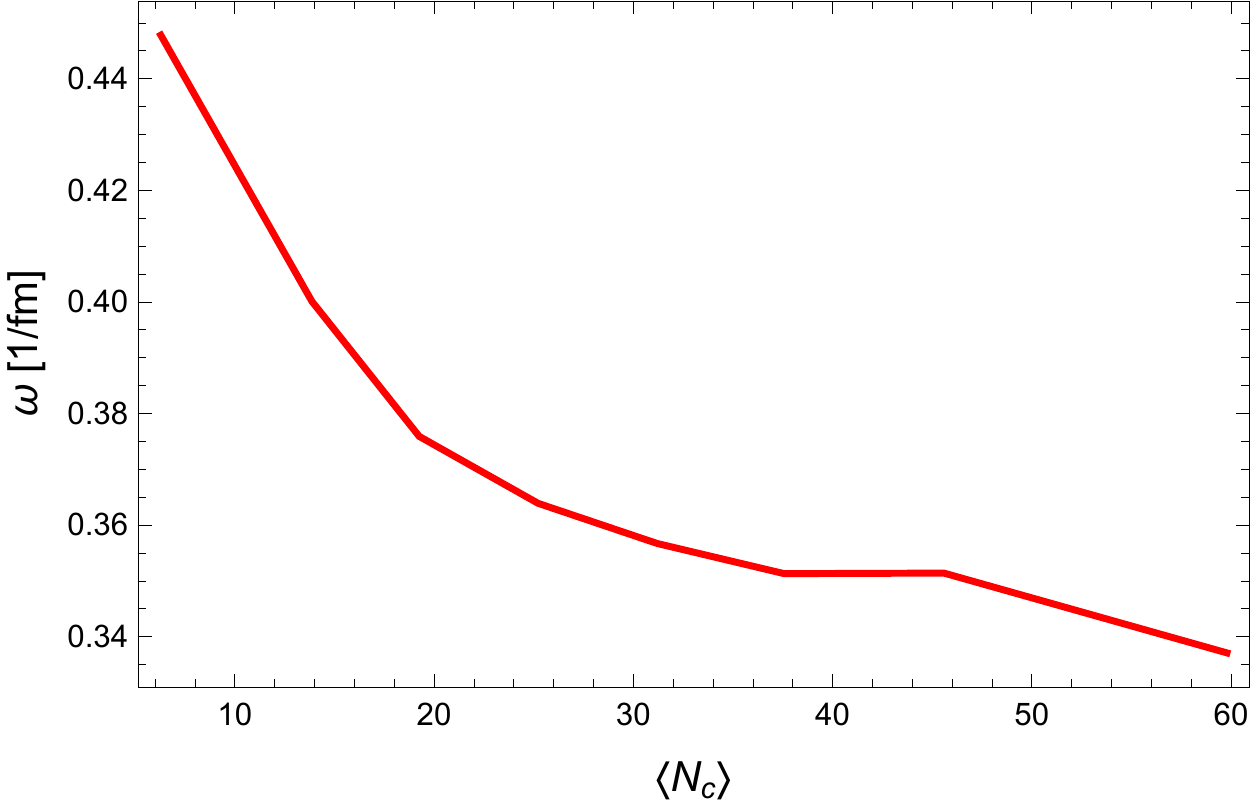}  \quad
\includegraphics[scale=0.5]{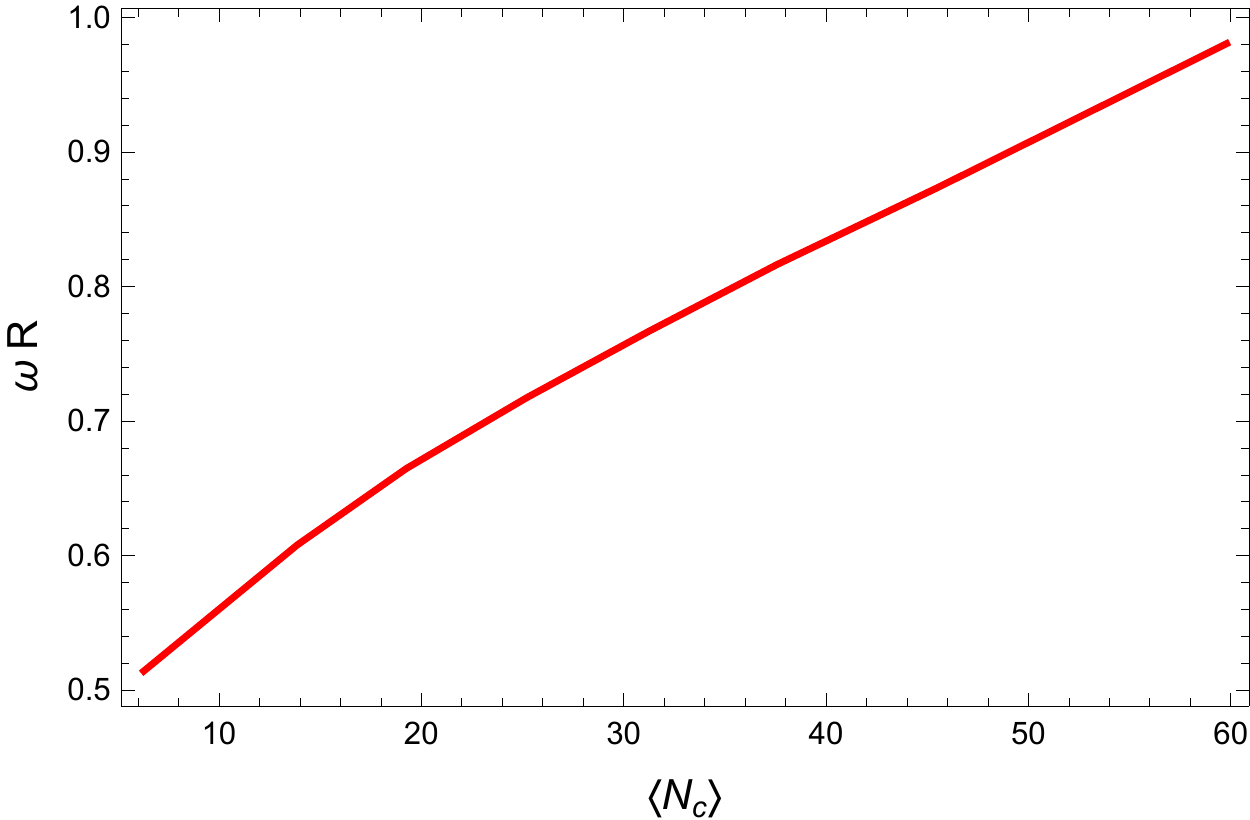} 
\end{center}
\caption{(Color online) The calculated flow parameters $\omega$ (left) and $\omega R$ (right) plotted vs. $N_c$.}%
\label{pt2}%
\end{figure}

The detailed comparison with data is given in the Appendix, where the the radii evaluated from the model and those measured by the ALICE collaborations are presented in Tables 3--10. Using these results and  formula (\ref{omegaRx}) one can evaluate the Hubble parameter $\omega$, responsible for the  strength of the radial flow. In Fig.~\ref{pt2} we show $\omega$ and $\omega R$  plotted vs. multiplicity $N_c$. One sees that the effect of the flow (as measured by $\omega R$) is non-negligible even at smallest multiplicities and increases substantially with increasing $N_c$ although the $\omega$
itself decreases.

\subsection {Correlation functions for $q  \neq 0$ }
\label{sect:corneq0}

We evaluate the HBT radii using (\ref{rform}) and consequently they are sensitive to the $q^2$ dependence   of the correlation functions only at very small $q$.   It is, however, necessary to verify if  the model does not give a clearly wrong behaviour of the correlation functions at larger values of $q$. We thus evaluated the correlation functions themselves in the region of $q$ up to $800$ MeV. It turns out that  for $q\geq 300$ MeV both $C_{\rm out}$ and $C_{\rm side}$ are rather sensitive to the value of $\D$, the relative width of the "shell" from which the particles are emitted. At  $\D <0.25$,  $C_{out}$  falls too slowly with increasing $q$ and  $C_{\rm side}$ shows large oscillations. Therefore  $\D$ must be restricted from below and thus  one cannot take too small  $\d$ (although the fit to the radii becomes even better for small $\d$).

\begin{figure}[t]
\begin{center}
\includegraphics[scale=0.8]{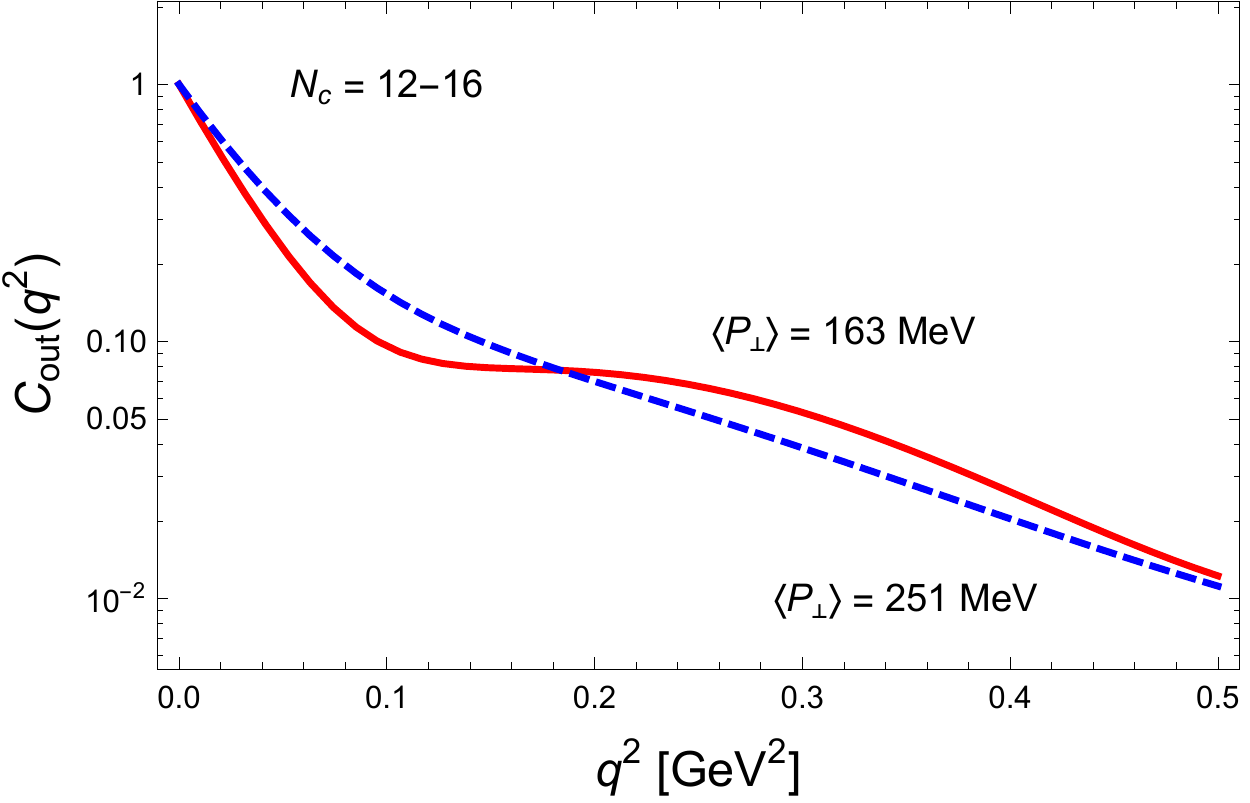}
\end{center}
\caption{(Color online) The function $C_{out} $ for the multiplicity bin $N_c$ = 12--16, plotted vs $q^2$, to illustrate deviations from the Gaussian behaviour. Full line: $\langle P_\p \rangle$=163 MeV. Dashed line: $\langle P_\p \rangle$=251 MeV. 
Note that the scale on the vertical axis is logarithmic, so that Gaussians would correspond to straight lines. }%
\label{pt1}%
\end{figure}

Using  the parameters as explained above, we evaluated the correlation functions for various multiplicities and transverse momenta. They look reasonable, except  at smallest $P_\p$, where the $C_{\rm out}$'s  exhibit  heavy tails and thus differ substantially from  Gaussians. This,  naturally, may influence the experimentally fitted HBT radii. This is illustrated in Fig. \ref{pt1} where   $C_{\rm out}(q)$ for the second multiplicity bin ($N_c$=12--16) is plotted vs.  $q^2$. One sees a rather dramatic difference between   $C_{\rm out}(q)$ at $\langle P_\p \rangle =163$ MeV and at $\langle P_\p \rangle =251$~MeV. It is clear from this figure that at $\langle P_\p \rangle =163$ MeV the fit to a Gaussian cannot provide a reliable value of the  $R_{\rm out}$.  For the first multiplicity bin ($N_c$~=~1--11) the effect is even stronger, while for the third ($N_c$ = 17--22) it is significantly weaker.  We feel that this may be  a possible explanation of the discrepancy of our model with data at this smallest $P_\p$.  

At larger  values of $\langle P_\p \rangle$ the deviations from Gaussians are important mostly in the region where the correlation functions are already rather small, and thus the effect seems to be contained within the (rather large) systematic errors quoted in~\cite{Aamodt:2011kd}. 

\subsection {The volume}

The data were taken for $|\e| \leq 1.2  \equiv \Delta\eta/2$. This allows us to evaluate the effective volume from which particles are emitted from the formula
\ba
V= \frac{2\pi \tau_f  \Delta \eta \,\int r dr f(r) \sqrt{1+\o^2 r^2}}{ f(r_s)\sqrt{1+\o^2 r_s^2}},
\ea
where $r_s$ is the point at which the function $  f(r) \sqrt{1+\o^2 r^2}$ takes the maximal value. The numerical evaluation of $V$ gives the values shown in Table~\ref{tab:vol}. The second line in Table~\ref{tab:vol}  gives the radius of the sphere of volume $V/N_c$. One sees that at larger multiplicities the pions are somewhat more tightly packed. 

\begin{table}[h]
{\begin{tabular}{ccccccccc} 
\hline \\
mult. class & 1--11 &  12--16  &  17--22 &  23--28 & 29--34 &  35--41 & 42--51 & 52--151 \\ 
\hline \\
$V \,\, [\hbox{fm}^3]$  &    46.2&  68.9&  86.8&  101.7&  116.6& 130.7& 142.9&  183.2 \\
$r_\pi$ [fm] & 1.21&  1.06&    1.02&   0.99&   0.96&  0.94&  0.91&  0.90 \\
\end{tabular}
}
\caption{The volume of the system at freeze-out for different multiplicity classes.}
\label{tab:vol}
\end{table}

\begin{figure}[h]
\begin{center}
\includegraphics[scale=0.75]{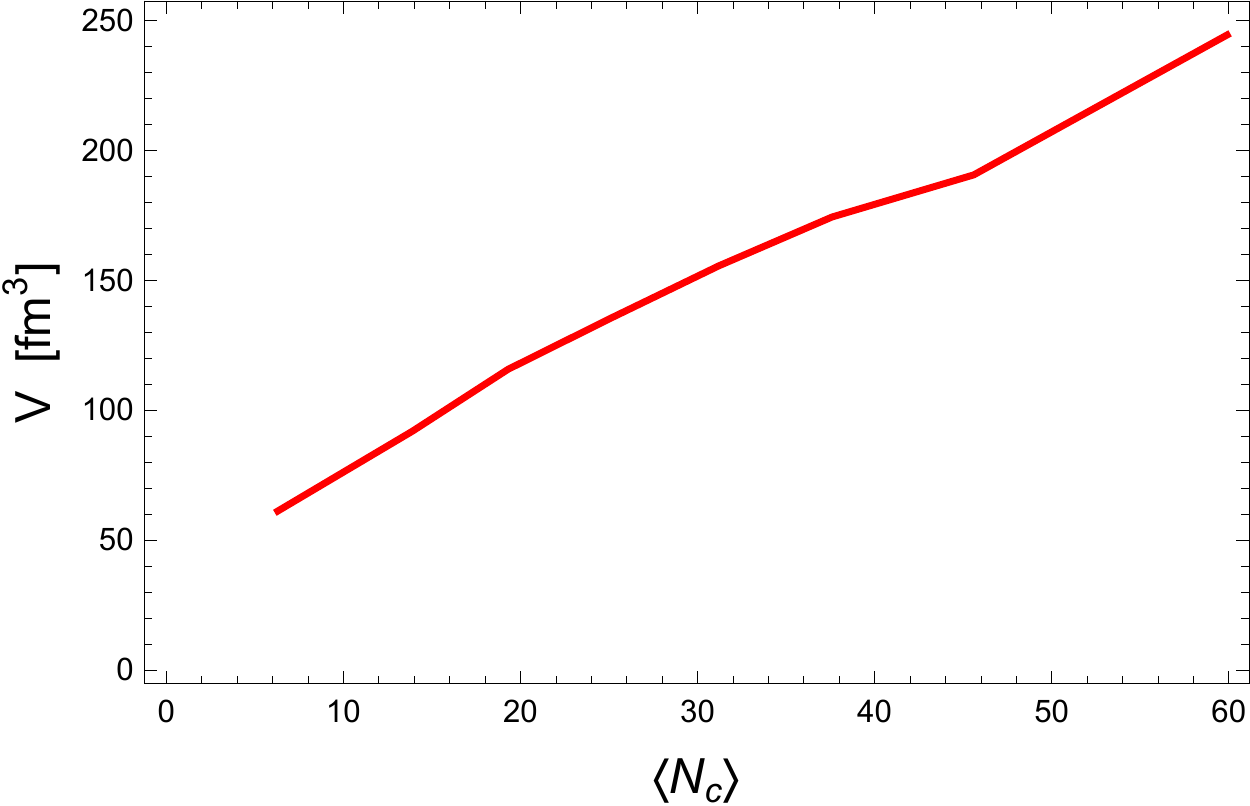} 
\end{center}
\caption{(Color online) Effective volume of the system as  function of the multiplicity~$N_c$.}%
\label{V}
\end{figure}

The graphical representation of the dependence of the volume $V$ on the mean multiplicity is shown in Fig.~\ref{V}.  We note that $V$ is much larger  than the product of the three HBT radii (at a given multiplicity). This is an expected result, the HBT radii measure the homogeneity lengths of the system rather than its physical dimensions \cite{Makhlin:1987gm}. The former are typically smaller than the latter. 

\section {Summary and conclusions}
\label{sect:sum+concl}

The main conclusion from our work is that the boost-invariant and azimuthally symmetric blast-wave model --- with a suitably selected transverse profile --- can account for the HBT radii in $pp$ collisions at 7 TeV measured by the ALICE collaboration. In particular, it has been possible to explain (i)  the general decrease of the HBT radii with increasing transverse momentum of the pair and (ii) the so-called {\it HBT puzzle}, i.e., small values of the ratio $R_{\rm out}/R_{\rm side}$. 

The blast-wave model, realizing a simple picture of a thermalised and expanding medium, allows to determine the relevant physical parameters describing the state of the system at the kinematic freeze-out (as measured by the Bose-Einstein correlations). Our analysis shows that the relevant temperature must be rather low (below 120 MeV). To obtain reasonable agreement with data, it was also necessary to introduce a certain amount of  transverse flow, which we assumed to be of the Hubble type (c.f. Eq.~(\ref{hub})).

The results concerning the geometry of the system at the (kinetic) freeze-out seem also  interesting. The first observation is that the transverse profile is far from a Gaussian: it rather resembles a shell which can be chosen to be of an approximately  constant width of
about  1.5--1.7 fm. This feature (when combined with the transverse flow) makes it possible  to explain the small ratio $R_{\rm out}/R_{\rm side}$. The radius of the shell increases steadily with increasing multiplicity (from $\sim$ 1 fm to $\sim$ 3 fm). Similarly, the proper time $\t$ at which the freeze-out takes place increases with the multiplicity from $\sim$ 2 fm to $\sim$ 3 fm. The increase is not linear but tends to saturate somewhat at high multiplicities.

It is interesting to note that a qualitatively similar shape of the transverse profile  was obtained by T. Csorgo \cite{csorgo1} in the analysis of the NA22 data on collisions of 250~GeV pions and kaons on the fixed proton target \cite{NA22}.  Thus the effect we observe seems to be a robust feature of hadron-hadron collisions, as it survived the change of almost three orders of magnitude in the c.m. collision energy.

It should be realized that the formulae we use to determine the HBT radii  describe the slope of the correlation function at $q^2=0$. In experiment, however, the radii are measured by fitting the observed correlation function to a Gaussian. These two procedures give identical results only if the correlation functions are indeed Gaussians.  In our calculations the non-gaussian shape of the transverse profile implies, of course,  deviations  of the measured correlation functions from the simple gaussian shape. We have found that this effect is not significant  for $R_{\rm long}$. For $C_{\rm out}$, however, it is essential at small $P_\p$. This may perhaps explain the anomalous behavior of the measured $R_{\rm out}$  in this region. 

We have shown that the blast-wave model provides a useful parametrization of the data on HBT radii measured by the ALICE collaboration in $pp$ collisions. This is achieved in terms of, basically, two parameters: the radius $R$ of the system and its (proper) life-time $\t$, both showing a regular dependence on the multiplicity. This result provides further support for the suggestion \cite{Ghosh:2014eqa}  that the thermalised, collectively expanding medium is formed even in such a small system as that created in $pp$ collisions.  

Let us add that the model provides also a host of predictions for the correlation functions at  large $q$. Detailed comparison  with the existing (although not yet published) ALICE data and with the future CMS data \cite{padula} in this $q$ region will provide a very strong test of the idea that the blast wave model can be applied even in $pp$ collisions.

 \vspace{0.3cm}
 
{\bf ACKNOWLEDGEMENTS}

We would like to thank Adam Kisiel for useful discussions and for  help in understanding the ALICE data and  Tamas Csorgo for pointing out to us the  results of NA22 experiment. This investigation was supported in part by the NCN Grants UMO-2013/09/B/ST2/00497 and DEC-2012/05/B/ST2/02528.

\section {Appendix}

In the Tables \ref{tab:111}--\ref{tab:5299} we list the values of the HBT radii (in fm) following from our fit, compared to those measured by the ALICE collaboration \cite{Aamodt:2011kd}. The first column: the mean transverse momentum of the pion pair, $P_\perp$; the second and third columns: $R_{\rm long}$ from the model calculation and the experiment, respectively; the fourth and fifth columns: $R_{\rm side}$; the sixth and seventh columns: $R_{\rm out}$. The listed errors represent systematic and statistical errors added in quadrature. The quality of the fits is shown in Fig.~\ref{radii} where the lines represent our model predictions, the central points of the bands correspond to the experimental results, and the width of the bands describes the experimental error.

\bigskip

\begin{table}[h]
{\begin{tabular}{ccccccc} 
Mult. class & 1--11 & & & & & \\
\hline \\
$P_\perp$ [GeV]
&  $R_{\rm long}$[fm] &  $R_{\rm long}$[fm]
&  $R_{\rm side}$[fm] &  $R_{\rm side}$[fm]
&  $R_{\rm out}$[fm] &  $R_{\rm out}$[fm]  \\ 
&  model &  exp.
&  model &  exp.
&  model &  exp. \\ 
\hline \\
0.163  & 1.54 & 1.58$\pm$0.37 &  0.75 & 0.75$\pm$0.10  & 0.95 & 0.49$\pm$0.10 \\
0.251  & 1.25 & 1.23$\pm$0.16 &  0.74 &0.76$\pm$0.10 &   0.87 & 0.79$\pm$0.13 \\
0.348 &  1.05 & 1.01$\pm$0.10  & 0.72 &0.73$\pm$0.10 &  0.77 & 0.78$\pm$0.11 \\
0.447  & 0.91 & 0.91$\pm$0.10  & 0.69 & 0.69$\pm$0.10 &  0.69 & 0.69$\pm$0.14 \\
0.547 &  0.80 & 0.85$\pm$0.10  & 0.66 & 0.66$\pm$0.10 &  0.63 & 0.64$\pm$0.15 \\
0.647 &  0.73 & 0.80$\pm$0.19 &  0.63 & 0.62$\pm$0.11 &  0.59 & 0.57$\pm$0.23 \\

\end{tabular}
}
\caption{Model results for the HBT radii compared with the experimental results for the multiplicity class $N$=1--11.}
\label{tab:111}
\end{table}

\begin{table}
{\begin{tabular}{ccccccc} 
Mult. class & 12--16 & & & & & \\
\hline \\
$P_\perp$ [GeV]
&  $R_{\rm long}$[fm] &  $R_{\rm long}$[fm]
&  $R_{\rm side}$[fm] &  $R_{\rm side}$[fm]
&  $R_{\rm out}$[fm] &  $R_{\rm out}$[fm]  \\ 
&  model &  exp.
&  model &  exp.
&  model &  exp. \\ 
\hline \\
0.163 &  1.74 & 1.80$\pm$0.29 &   0.98 & 1.06$\pm$0.11 &   1.13 & 0.78$\pm$0.12 \\
0.251 &  1.41 & 1.38$\pm$0.13 &   0.94 & 0.97$\pm$0.10 &  1.00 & 1.01$\pm$0.11 \\
0.348 & 1.18 & 1.15$\pm$0.10 &   0.89 & 0.89$\pm$0.10 & 0.85 & 0.84$\pm$0.10 \\
0.448 & 1.02 & 1.02$\pm$0.10 & 0.84 & 0.83$\pm$0.10 &   0.74 & 0.73$\pm$0.10 \\
0.547 & 0.91 & 0.95$\pm$0.10 &  0.79 & 0.78$\pm$0.10 &   0.67 & 0.61$\pm$0.10 \\
0.647 &   0.83 & 0.96$\pm$0.15 &   0.75 & 0.77$\pm$0.10 &   0.62 & 0.54$\pm$0.14
\end{tabular}
}
\caption{Same as Table \ref{tab:111} but for the multiplicity class $N$=12--16.}
\label{tab:1216}
\end{table}

\begin{table}
{\begin{tabular}{ccccccc} 
Mult. class & 17--22 & & & & & \\
\hline \\
$P_\perp$ [GeV]
&  $R_{\rm long}$[fm] &  $R_{\rm long}$[fm]
&  $R_{\rm side}$[fm] &  $R_{\rm side}$[fm]
&  $R_{\rm out}$[fm] &  $R_{\rm out}$[fm]  \\ 
&  model &  exp.
&  model &  exp.
&  model &  exp. \\ 
\hline \\
0.163 & 1.86 & 1.88$\pm$0.30  & 1.14 & 1.18$\pm$0.11 & 1.24 & 0.99$\pm$0.14 \\
0.251 & 1.52 & 1.49$\pm$0.14  & 1.08 & 1.11$\pm$0.10 & 1.07 & 1.11$\pm$0.12 \\
0.349 & 1.27 & 1.22$\pm$0.11  & 1.00 & 0.98$\pm$0.10 & 0.90 & 0.90$\pm$0.10 \\
0.448 & 1.10 & 1.12$\pm$0.11  & 0.94 & 0.90$\pm$0.10 & 0.77 & 0.72$\pm$0.10 \\
0.548 & 0.98 & 1.03$\pm$0.10  & 0.88 & 0.87$\pm$0.10 & 0.69 & 0.62$\pm$0.10 \\
0.647 & 0.89 & 1.00$\pm$0.16  & 0.83 & 0.89$\pm$0.10 & 0.64 & 0.54$\pm$0.14
\end{tabular}
}
\caption{Same as Table \ref{tab:111} but for the multiplicity class $N$=17--22.}
\label{tab:1722}
\end{table}

\begin{table}
{\begin{tabular}{ccccccc} 
Mult. class & 23--28 & & & & & \\
\hline \\
$P_\perp$ [GeV]
&  $R_{\rm long}$[fm] &  $R_{\rm long}$[fm]
&  $R_{\rm side}$[fm] &  $R_{\rm side}$[fm]
&  $R_{\rm out}$[fm] &  $R_{\rm out}$[fm]  \\ 
&  model &  exp.
&  model &  exp.
&  model &  exp. \\ 
\hline \\
0.163 & 1.94 & 1.99$\pm$0.31 & 1.26 & 1.30$\pm$0.12 & 1.32 & 1.18$\pm$0.17 \\
0.251 & 1.58 & 1.56$\pm$0.15 & 1.18 & 1.21$\pm$0.10 & 1.12 & 1.15$\pm$0.13 \\
0.349 & 1.33 & 1.29$\pm$0.11 & 1.09 & 1.06$\pm$0.10 & 0.92 & 0.93$\pm$0.10 \\
0.448 & 1.15 & 1.15$\pm$0.11 & 1.01 & 0.99$\pm$0.10 & 0.79 & 0.73$\pm$0.10 \\
0.548 & 1.03 & 1.05$\pm$0.11 & 0.94 & 0.97$\pm$0.10 & 0.70 & 0.63$\pm$0.10 \\
0.648 & 0.93 & 1.13$\pm$0.19 & 0.89 & 0.91$\pm$0.12 & 0.65 & 0.48$\pm$0.13
\end{tabular}
}
\caption{Same as Table \ref{tab:111} but for the multiplicity class $N$=23--28.}
\label{tab:2328}
\end{table}

\begin{table}
{\begin{tabular}{ccccccc} 
Mult. class & 29--34 & & & & & \\
\hline \\
$P_\perp$ [GeV]
&  $R_{\rm long}$[fm] &  $R_{\rm long}$[fm]
&  $R_{\rm side}$[fm] &  $R_{\rm side}$[fm]
&  $R_{\rm out}$[fm] &  $R_{\rm out}$[fm]  \\ 
&  model &  exp.
&  model &  exp.
&  model &  exp. \\ 
\hline \\
0.163 & 2.01 & 1.98$\pm$0.31 & 1.36 & 1.35$\pm$0.12 & 1.38 & 1.23$\pm$0.17 \\
0.251 & 1.64 & 1.60$\pm$0.15 & 1.26 & 1.30$\pm$0.10 & 1.15 & 1.20$\pm$0.14 \\
0.349 & 1.37 & 1.32$\pm$0.11 & 1.16 & 1.09$\pm$0.11 & 0.94 & 0.90$\pm$0.10 \\
0.448 & 1.20 & 1.16$\pm$0.11 & 1.07 & 1.06$\pm$0.11 & 0.80 & 0.75$\pm$0.10 \\
0.548 & 1.06 & 1.18$\pm$0.11 & 1.00 & 1.01$\pm$0.11 & 0.71 & 0.61$\pm$0.10 \\
0.648 & 0.97 & 1.13$\pm$0.19 & 0.93 & 1.05$\pm$0.13 & 0.65 & 0.52$\pm$0.15
\end{tabular}
}
\caption{Same as Table \ref{tab:111} but for the multiplicity class $N$=29--34.}
\label{tab:2934}
\end{table}

\begin{table}
{\begin{tabular}{ccccccc} 
Mult. class & 35--41 & & & & & \\
\hline \\
$P_\perp$ [GeV]
&  $R_{\rm long}$[fm] &  $R_{\rm long}$[fm]
&  $R_{\rm side}$[fm] &  $R_{\rm side}$[fm]
&  $R_{\rm out}$[fm] &  $R_{\rm out}$[fm]  \\ 
&  model &  exp.
&  model &  exp.
&  model &  exp. \\ 
\hline \\
0.163 & 2.06 & 1.99$\pm$0.31 & 1.46 & 1.43$\pm$0.12 & 1.44 & 1.34$\pm$0.19 \\
0.251 & 1.69 & 1.63$\pm$0.15 & 1.35 & 1.35$\pm$0.13 & 1.17 & 1.22$\pm$0.14 \\
0.349 & 1.41 & 1.37$\pm$0.11 & 1.23 & 1.17$\pm$0.11 & 0.95 & 0.92$\pm$0.11 \\
0.448 & 1.23 & 1.22$\pm$0.11 & 1.13 & 1.12$\pm$0.11 & 0.80 & 0.75$\pm$0.11 \\
0.548 & 1.10 & 1.19$\pm$0.11 & 1.04 & 1.07$\pm$0.11 & 0.71 & 0.60$\pm$0.10 \\
0.648 & 1.00 & 1.15$\pm$0.20 & 0.98 & 1.14$\pm$0.14 & 0.66 & 0.54$\pm$0.15
\end{tabular}
}
\caption{Same as Table \ref{tab:111} but for the multiplicity class $N$=35--41.}
\label{tab:3541}
\end{table}

\begin{table}
{\begin{tabular}{ccccccc} 
Mult. class & 42--51 & & & & & \\
\hline \\
$P_\perp$ [GeV]
&  $R_{\rm long}$[fm] &  $R_{\rm long}$[fm]
&  $R_{\rm side}$[fm] &  $R_{\rm side}$[fm]
&  $R_{\rm out}$[fm] &  $R_{\rm out}$[fm]  \\ 
&  model &  exp.
&  model &  exp.
&  model &  exp. \\ 
\hline \\
0.163 & 2.07 & 2.14$\pm$0.34 & 1.55 & 1.53$\pm$0.13 & 1.48 & 1.43$\pm$0.21 \\
0.251 & 1.70 & 1.66$\pm$0.16 & 1.41 & 1.41$\pm$0.13 & 1.18 & 1.24$\pm$0.14 \\
0.349 & 1.42 & 1.33$\pm$0.11 & 1.28 & 1.21$\pm$0.11 & 0.94 & 0.94$\pm$0.11 \\
0.448 & 1.24 & 1.23$\pm$0.11 & 1.17 & 1.12$\pm$0.11 & 0.80 & 0.69$\pm$0.11 \\
0.548 & 1.10 & 1.23$\pm$0.11 & 1.08 & 1.18$\pm$0.11 & 0.71 & 0.61$\pm$0.11 \\
0.648 & 1.00 & 1.27$\pm$0.22 & 1.01 & 1.25$\pm$0.17 & 0.66 & 0.42$\pm$0.15
\end{tabular}
}
\caption{Same as Table \ref{tab:111} but for the multiplicity class $N$=42--51.}
\label{tab:4251}
\end{table}
\begin{table}
{\begin{tabular}{ccccccc} 
Mult. class & 52--151 & & & & & \\
\hline \\
$P_\perp$ [GeV]
&  $R_{\rm long}$[fm] &  $R_{\rm long}$[fm]
&  $R_{\rm side}$[fm] &  $R_{\rm side}$[fm]
&  $R_{\rm out}$[fm] &  $R_{\rm out}$[fm]  \\ 
&  model &  exp.
&  model &  exp.
&  model &  exp. \\ 
\hline \\
0.163 & 2.21 & 2.14$\pm$0.50 & 1.78 & 1.63$\pm$0.19 & 1.59 & 1.60$\pm$0.32 \\
0.251 & 1.81 & 1.77$\pm$0.24 & 1.60 & 1.54$\pm$0.21 & 1.23 & 1.28$\pm$0.22 \\
0.349 & 1.52 & 1.49$\pm$0.14 & 1.43 & 1.39$\pm$0.14 & 0.97 & 0.97$\pm$0.14 \\
0.449 & 1.33 & 1.25$\pm$0.14 & 1.30 & 1.27$\pm$0.14 & 0.81 & 0.77$\pm$0.16 \\
0.548 & 1.18 & 1.39$\pm$0.17 & 1.20 & 1.33$\pm$0.15 & 0.73 & 0.62$\pm$0.16 \\
0.648 & 1.08 & 1.32$\pm$0.33 & 1.12 & 1.29$\pm$0.24 & 0.67 & 0.35$\pm$0.16
\end{tabular}
}
\caption{Same as Table \ref{tab:111} but for the multiplicity class $N$=52--151.}
\label{tab:5299}
\end{table}

\FloatBarrier

\end{document}